\documentclass[showpacs, amsmath, amssymb, onecolumn, pra, aps, superscriptaddress, notitlepage, twocolumn, longbibliography]{revtex4-2}

\usepackage{amsmath, amssymb, amsthm, mathrsfs, amsfonts, dsfont}
\usepackage[dvips]{graphicx}
\usepackage{subcaption}
\usepackage{braket}
\usepackage{bm}
\usepackage{enumerate}
\usepackage{color}
\usepackage{comment}
\usepackage[normalem]{ulem}
\usepackage{booktabs}
\usepackage{siunitx}
\usepackage[whole]{bxcjkjatype}
\usepackage{url}
\usepackage{hyperref}
\usepackage{qcircuit}
\usepackage{makecell}

\newtheorem*{theorem*}{Theorem}

\hypersetup{colorlinks, citecolor=blue, linkcolor=blue, urlcolor=blue}

\newcommand{\ceil}[1]{\left\lceil#1\right\rceil}

\begin{document}
\title{Network-Based Quantum Computing: an efficient design framework for many-small-node distributed fault-tolerant quantum computing}

\author{Soshun Naito}
\email{naito@biom.t.u-tokyo.ac.jp}
\affiliation{Department of Information and Communication Engineering, Graduate School of Information Science and Technology, The University of Tokyo, 7-3-1 Hongo, Bunkyo-ku, Tokyo 113-0033, Japan}

\author{Yasunari Suzuki}
\email{yasunari.suzuki@riken.jp}
\affiliation{NTT Computer and Data Science Laboratories, NTT Inc., Musashino 180-8585, Japan} 
\affiliation{Center for Quantum Computing, RIKEN, 2-1 Hirosawa, Wako, Saitama 351-0198, Japan}

\author{Yuuki Tokunaga}
\affiliation{NTT Computer and Data Science Laboratories, NTT Inc., Musashino 180-8585, Japan}

\begin{abstract}
    In fault-tolerant quantum computing, a large number of physical qubits are required to construct a single logical qubit, and a single quantum node may be able to hold only a small number of logical qubits. In such a case, the idea of distributed fault-tolerant quantum computing~(DFTQC) is important to demonstrate large-scale quantum computation using small-scale nodes. However, the design of distributed systems on small-scale nodes, where each node can store only one or a few logical qubits for computation, has not been explored well yet. In this paper, we propose network-based quantum computation~(NBQC) to efficiently realize distributed fault-tolerant quantum computation using many small-scale nodes. A key idea of NBQC is to let computational data continuously move throughout the network while maintaining the connectivity to other nodes. We numerically show that, for practical benchmark tasks, our method achieves shorter execution times than circuit-based strategies and more node-efficient constructions than measurement-based quantum computing. Also, if we are allowed to specialize the network to the structure of quantum programs, such as peak access frequencies, the number of nodes can be significantly reduced.  Thus, our methods provide a foundation in designing DFTQC architecture exploiting the redundancy of many small fault-tolerant nodes.
\end{abstract}

\maketitle

\section{Introduction}
One of the most challenging obstacles to realizing utility-scale fault-tolerant quantum computing~(FTQC) is scaling up FTQCs to a meaningful size without compromising error rates and control speed. This gap is significant, as meaningful quantum applications would require at least tens of logical qubits implemented using a few tens of thousands of physical qubits~\cite{beverland2022assessing,yoshioka2024hunting}, while several hardware restrictions limit the number of qubits available per node~\cite{nagayama2017surface,hertzberg2021laser,strikis2023quantum,siegel2023adaptive}.
A promising approach to overcome this limitation is distributed fault-tolerant quantum computation (DFTQC), in which multiple fault-tolerant quantum nodes containing a number of logical qubits are connected by fault-tolerant quantum channels to form a larger FTQC~\cite{van2010distributed,nickerson2014freely}. In DFTQC, quantum programs consist of local and remote logical operations, where each remote operation consumes a distilled logical Bell state shared between nodes. Since the latency for generating logical Bell states is typically much longer than that of local operations~\cite{pattison2025constant,maeda2025logical}, it is crucial to mitigate the overhead by slow quantum communication for efficient DFTQC execution.  

This paper focuses on the many-small-node regime of DFTQC, where each node contains a small number of logical qubits and can communicate with only a few other nodes through slow logical channels (see Fig.~\ref{fig:positioning} and Table.~\ref{tab:positioning}). We assume one or a few logical qubits per node can be used for storing computational data, named algorithmic qubit, since logical operations and fault-tolerant Bell-state distillation also require a number of ancillary logical qubits~\cite{beverland2022assessing,fowler2018low,pattison2025constant}. 

Massive efforts have been devoted to developing efficient DFTQC architectures. Existing proposals can be categorized into two paradigms: \textit{circuit-based DFTQC} (CB-DFTQC) and \textit{measurement-based DFTQC} (MB-DFTQC).  
The circuit-based paradigm constructs an FTQC network by assigning each algorithmic qubit to a specific node and executing quantum programs by converting several operations into remote ones. While this approach is conceptually simple, a large fraction of instructions become remote operations in the many-small-node regime, resulting in significant communication overheads. Such overheads can be alleviated through optimizations of network-structure design, circuit partitioning, and instruction scheduling~\cite{andres2019automated,zomorodi2018optimizing,wu2022autocomm,zhang2024optimizing,main2025distributed,daei2020optimized,nikahd2021automated,burt2024generalised,andres2024distributing,sundaram2022distribution,sundaram2023distributing,cambiucci2023hypergraphic,baker2020time,ovide2023mapping,bandic2023mapping,pastor2024circuit}. However, it remains inherently difficult to eliminate communication latency.  
The other paradigm leverages the framework of measurement-based quantum computation (MBQC)~\cite{raussendorf2001one,raussendorf2003measurement,broadbent2009universal}. MBQC executes quantum programs by first generating a program-independent entangled cluster state and then performing program-dependent adaptive local measurements. In measurement-based DFTQC within the many-small-node regime, a large cluster state can be envisioned in which each node corresponds to a fault-tolerant node with logical qubits.
An advantage of MBQC is that it is intrinsically free from repeated communication overheads, since logical Bell states between nodes are generated only once at the beginning; thus, MB-DFTQC can achieve an execution time comparable to that of a single-node FTQC, i.e., FTQC on one large fault-tolerant quantum node. However, MB-DFTQC requires an enormous number of nodes, and its node-count optimization is non-trivial. Also, it is ambiguous how to treat probabilistic processes such as magic-state generations in this framework. These issues prevent us from employing MBQC as a standard DFTQC architecture. Therefore, to achieve efficient DFTQC in the early stages of FTQC development, it is essential to establish a new paradigm that combines high performance with moderate resource requirements of these approaches.

Here, we propose a fast and node-count-efficient paradigm for DFTQC with many small nodes, called \textit{Network-Based Quantum Computing~(NBQC)}. The key idea of NBQC is to combine the node efficiency of CB-DFTQC with the overhead-concealing mechanism of MB-DFTQC. In NBQC, algorithmic qubits are not fixed to specific nodes but continuously move throughout the network to maintain connectivity with all other algorithmic qubits. Although this design introduces additional nodes, NBQC reduces overall latency with moderate node-count overhead by employing efficient network components, including a ring network, a switching network, components for magic-state generation, and inter-component communication links~(see Fig.~\ref{fig:NBQC_overview}).
Owing to this architecture, NBQC can achieve an execution time nearly identical to that of MB-DFTQC while accounting for magic-state generation, i.e., it can hide most of the communication latency except during the initial Bell-state generation. Moreover, if we can tailor the network to the access patterns of remote operations (explained in the main text) in typical target applications, NBQC can further reduce the required number of nodes without introducing communication overhead. Even when the number of available nodes is limited but redundant, NBQC can effectively exploit such redundancy to conceal communication overhead. We also propose several heuristic protocols to suppress the total number of nodes in NBQC architectures.  

We numerically evaluated NBQC in terms of node counts and execution times, comparing them against CB-DFTQC and MB-DFTQC. Our results show that, given sufficient nodes, NBQC accounting magic-state generation achieves an execution time and node count nearly identical to those of MB-DFTQC ignoring magic-state generation. If we are allowed to assume access patterns, NBQC achieves significantly fewer nodes compared to MB-DFTQC and much shorter execution times than CB-DFTQC across practical benchmarks. By tuning the number of available nodes, NBQC achieves a smooth trade-off between node count and execution time. 

While our numerical simulations assume each node stores a single logical qubit encoded with surface codes, Sec.\,\ref{sec:discussion} discusses that NBQC can be applied to more general cases, such as non-reconfigurable networks, non-2D architectures such as neutral atoms, or quantum low-density parity-check codes encoding multiple logical qubits.
Therefore, we believe that NBQC provides a practical and scalable foundation for constructing distributed fault-tolerant quantum computers composed of many small nodes.  

Our contributions are summarized as follows.
\begin{itemize}
    \item We propose \textit{Network-Based Quantum Computing (NBQC)}, a DFTQC framework targeting many-small-node regimes. NBQC efficiently hides quantum communication latency by using components specifically designed for this regime. It serves as a hybrid paradigm that combines CB-DFTQC and MB-DFTQC, enabling fast fault-tolerant quantum computation with modest additional resource overhead.
    \item We develop a concrete protocol to generate an efficient NBQC network. To this end, we present an algorithm that reduces network size without compromising latency by identifying and removing unnecessary nodes. While this construction is tailored to a target circuit, the resulting network would exhibit similar performance for other circuits with a similar access pattern.
    \item We numerically evaluate NBQC on various benchmark tasks. As a result, NBQC achieves an algorithmic execution time while accounting for magic-state preparation. By assuming access patterns of remote operations, NBQC shows significantly fewer nodes than MB-DFTQC while maintaining comparable computation times, and achieves substantially shorter execution times than CB-DFTQC. NBQC also exhibits a smooth trade-off between node availability and execution latency.
\end{itemize}

This paper is organized as follows. We explain the background of DFTQCs, review existing approaches, and clarify the motivation in designing DFTQCs for a many-small-node regime in Sec.\,\ref{sec:preliminary}. Sec.\,\ref{sec:nbqc} presents our proposal, network-based quantum computing, and shows that it can conceal the communication overheads if a sufficient number of nodes are provided. Then, Sec.\,\ref{sec:optimize} shows the construction of the NBQC network according to the provided number of redundant nodes. The numerical evaluation for the performance of NBQC and its comparison to CB-DFTQC and MB-DFTQC are shown in Sec.\,\ref{sec:evaluation}. Sec.\,\ref{sec:discussion} discusses the extension of NBQC designs and their qualitative difference to the existing approaches. Finally, Sec.\,\ref{sec:conclusion} is devoted to a summary and conclusions.

\section{Preliminaries}
\label{sec:preliminary}
\subsection{Distributed Fault-Tolerant Quantum Computation}
For practical computational advantage, quantum computation needs a large number of qubits with sufficiently low error rates. Since physical qubits typically have high error rates, quantum error correction~(QEC) must be integrated to reduce logical error rates to an acceptable level. Even with state-of-the-art QEC codes, it is estimated that a few tens or hundreds of physical qubits are required to encode a single logical qubit~\cite{fowler2018low,bravyi2024high,xu2024constant,gidney2025yoked} to attain logical error rates sufficient for quantum advantage. 
On the other hand, recent experimental results report that the available physical qubit count per node, such as a superconducting circuit chip, single optical trap, or quantum charge-coupled device chip, is limited to approximately $10^{2}$ to $10^{4}$~\cite{google2023suppressing,google2025quantum,bluvstein2024logical,yoder2025tour} due to qubit imperfections such as crosstalk, yield, defects, and so on~\cite{hertzberg2021laser,nagayama2017surface,strikis2023quantum,siegel2023adaptive}. This means the number of logical qubits per node will be limited to around $10$ and is insufficient to demonstrate quantum advantage. Thus, scaling up these devices remains a substantial challenge.

One promising approach to addressing this limitation is using distributed fault-tolerant quantum computing~(DFTQC). In this approach, we perform quantum computing with several nodes that are interconnected via noisy quantum channels. While quantum channels also have high error rates, we can perform a high-fidelity logical remote CNOT gate and quantum state teleportation by consuming a high-fidelity entangled pair encoded in logical qubits. Here, each logical entanglement can be generated by distributing noisy physical Bell pairs, encoding them into logical qubits, and distilling noisy logical Bell pairs to high-fidelity ones~\cite{bennett1996mixed,ataides2025constant,pattison2025constant,maeda2025logical,ramette2024fault,shi2025stabilizer,leone2025resource}. Then, we can perform large-scale FTQC in a fault-tolerant manner across multiple nodes.

A major challenge in DFTQC is the degradation of execution time caused by limited quantum communication bandwidth. The number of quantum channels and the rate of physical entanglement generation are technologically limited. Since the generation of high-fidelity entangled pairs of logical qubits requires a massive number of physical entangled qubits and local operations, its generation rate is expected to be significantly slower than the speed of local logical operations. 
While the number of physical entanglements per logical entanglement can be reduced by using QEC codes with high encoding rates~\cite{bennett1996mixed}, they demand a few tens of logical qubits as a buffer space for efficient logical entanglement distillation~\cite{pattison2025constant}, which would be challenging for early devices.
If the generation rate of logical entangled pairs is insufficient, the execution time of the FTQC process is dominated by communication time, which makes it challenging to demonstrate quantum advantage.
Therefore, it is crucial to design high-performance DFTQC architectures under constraints of limited logical qubit capacity, limited quantum channels, and slow entangled pair generation rates.

\subsection{Focus of this work: DFTQCs with many small nodes}
\begin{figure*}[t]
    \centering
    \includegraphics[width=0.8\linewidth]{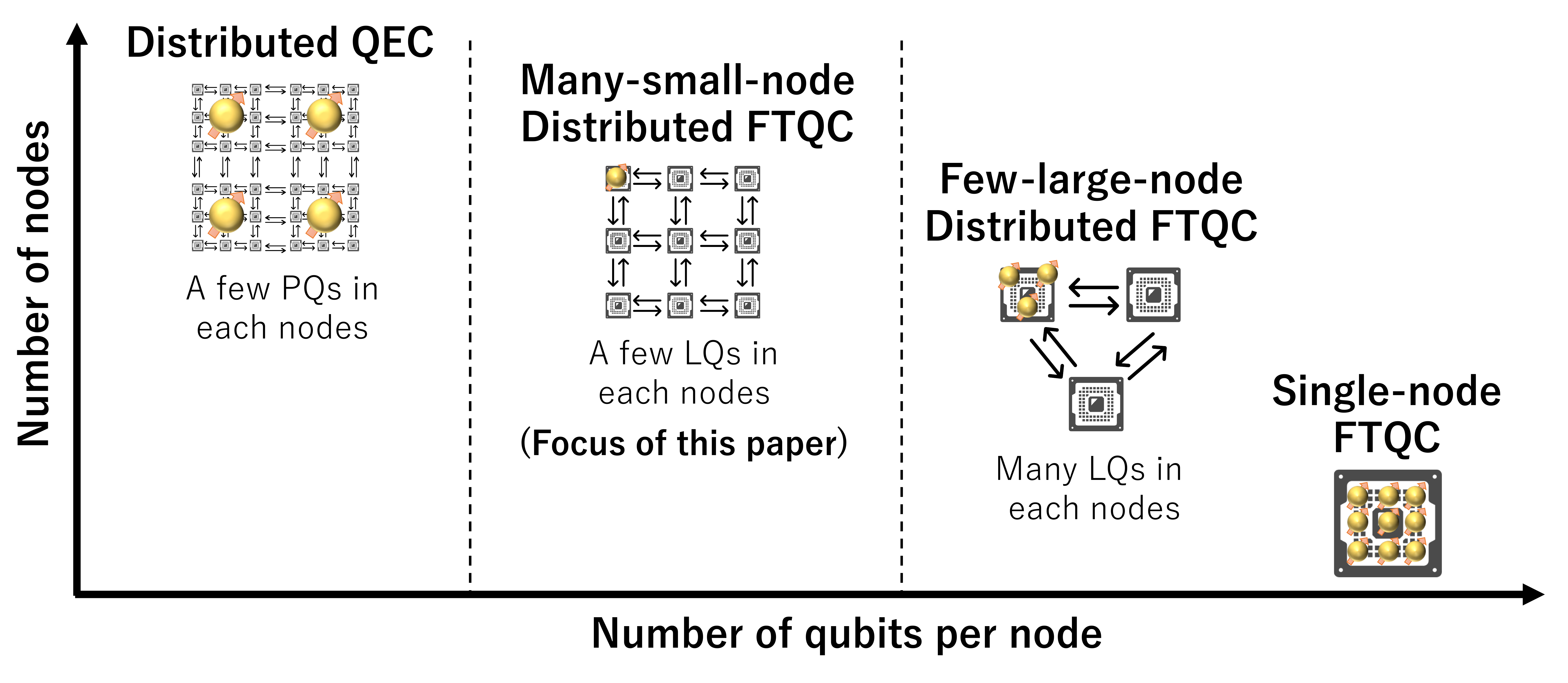}
    \caption{Design classification of DFTQC according to the number of nodes and the number of qubits per node. The left regime, distributed QEC, does not require integration technologies but demands quantum communication that is faster than the syndrome measurement cycle and has error rates lower than the code threshold, which is challenging for current technology. The right regime, few-large-node DFTQC or single-node FTQC, would be fast and simple, but its integration is challenging. In the middle regime, many-small-node DFTQC can be implemented with slow quantum communication and modest integration technology. Still, it might suffer from the execution time penalty due to massive communication. This paper focuses on the middle regime.}
    \label{fig:overview}
\end{figure*}

While the performance of each node might be technologically limited, there is room for increasing the number of nodes. Therefore, in this work, we investigate designs of DFTQCs with many small nodes. This regime involves many quantum nodes, each containing only a few logical qubits, and is connected by a few logical channels that are much slower than local logical operations. Table\,\ref{tab:positioning} and Figure\,\ref{fig:overview} summarize typical regimes of FTQC, the focus of this paper, and the requirement to demonstrate quantum advantage. This paper studies the potential of accelerating the DFTQC using redundant nodes. 

\begin{table*}[t]
    \caption{Regimes of DFTQCs according to the number of qubits per node. \texttt{\#PQ/node} and \texttt{\#LQ/node} represent the number of physical qubits and logical qubits per node, respectively. Here, $n_{\rm enc}$ is the number of physical qubits per logical qubit, and $n_{\rm alg}$ is the number of algorithmic qubits required for quantum advantage.
    According to the recent resource estimation and the comparison to the execution time of classical computing~\cite{yoshioka2024hunting,babbush2018encoding,gidney2021factor}, these values are approximately $n_{\rm enc} \sim 10^3$ and $n_{\rm alg} \sim 10^3$ if we suppose surface codes with $0.1\%$ physical error rates and early applications with exponential speed-up.}
    \label{tab:positioning}
    \begin{tabular}{|l||l|l|l|}
        \hline
        \texttt{\#PQ/node}        & \texttt{\#LQ/node}        & Regime & Requirement \\
        \hline
        \hline
        $ \sim 10$    & 0           & \makecell{Distributed QEC \\ or measurement-based FTQC} & \makecell{A single logical qubit is constructed across many nodes. \\ The reduction of logical error rates needs \\ fast, parallel, and high-fidelity physical communications.} \\
        \hline
        $ \sim 10 n_{\rm enc}$ & $\sim 10$    & \makecell{Many-small-node \\ Distributed FTQC \\ \textbf{(Focus of this paper)}} & \makecell{A few logical qubits are constructed within a node. \\ Most logical operations are non-local, \\ and remote operations can be a dominant bottleneck in execution time.} \\
        \hline
        $< n_{\rm alg} n_{\rm enc}$ & $< n_{\rm alg}$ & \makecell{Few-large-node \\ Distributed FTQC} & \makecell{Many logical qubits are constructed within a node. \\ Most logical operations are local, \\ and remote operations might not be a significant overhead.} \\
        \hline
        $\ge n_{\rm alg} n_{\rm enc}$ & $\ge n_{\rm alg}$ & \makecell{Single-node FTQC} & \makecell{All the logical qubits can be implemented within a node. \\ Applications can be executed with a single node, \\ and no remote logical operations are needed.} \\
        \hline
    \end{tabular}
\end{table*}

For clarity, we denote the parameters of DFTQCs as follows. Suppose a quantum circuit with $n_{\text{alg}}$ qubits and $D$ depth, which we call algorithmic qubit count and algorithmic depth, respectively. We aim to execute this quantum circuit on DFTQC as fast as possible.
We assume that DFTQC is composed of $N$ nodes, each of which can store $n_{\text{node}}$ algorithmic qubits for computation and can perform local logical operations with time $T_{\text{local}}$. We assume each node can establish at most $d$ logical channels to other nodes, and each channel generates a logical entangled pair in each period $T_{\text{Bell}}$. 
Note that ancillary logical qubit space for local logical operations, e.g., twist or lattice surgery~\cite{beverland2022assessing,fowler2018low}, and buffers for entanglement distillation~\cite{pattison2025constant} are prepared separately to $n_{\text{node}}$.

\begin{figure*}[t]
    \centering
    \includegraphics[width=0.9\linewidth]{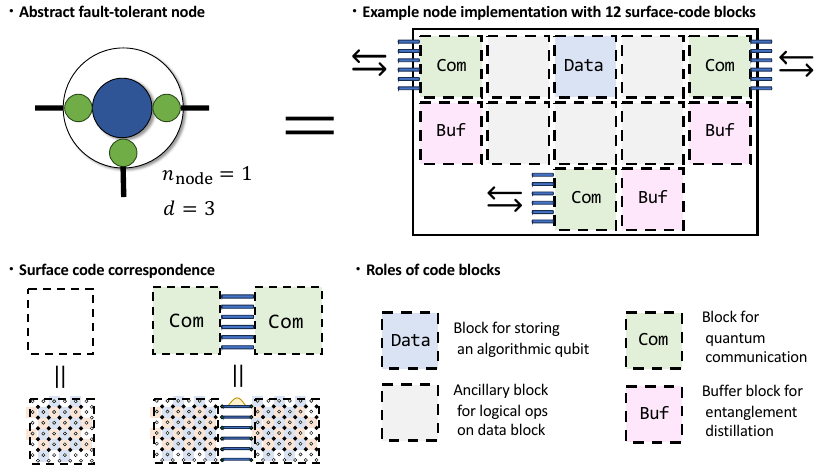}
    \caption{An example implementation of fault-tolerant nodes with $n_{\rm node}=1$ and $d=3$. Each node contains a single data code block~(blue cell), three sets of entanglement distillers, and ancillary qubits for local logical operations. Each communication code block~(green cell) has a quantum channel to that in other nodes, and can prepare logical entangled states with low fidelity. They can be distilled using buffer cells~(pink cell) to distill high-fidelity entanglement. Ancillary cells~(gray cell) are used for performing logical $H$ and $S$ gates on data block~\cite{fowler2018low,beverland2022assessing,brown2017poking} and local two-qubit gates to consume distilled entanglement or entanglement swapping. If each code block is implemented with the surface code, it corresponds to a 2D array of physical qubits, as shown in the bottom left. Here, blue and brown squares correspond to $X$ and $Z$ stabilizer measurements and circles to data qubits. Communication blocks have additional qubits that are physically connected to the other. See Sec.\,\ref{sec:discussion} for more general situations.}
    \label{fig:example_node_impl}
\end{figure*}

This work focuses on a many-small-node regime, i.e., $n_{\text{node}} = 1$, $d \sim 3$, and $T_{\text{Bell}} \gg T_{\text{local}}$, but we have sufficient number of nodes as $N \gg n_{\text{alg}} / n_{\text{node}}$. Note that we assume $n_{\text{node}} = 1$ to simplify the explanation of our proposals, and our proposal can be extended to cases for $n_{\text{node}} > 1$ as discussed in Sec.\,\ref{sec:discussion}. See Fig.\,\ref{fig:example_node_impl} for an example implementation of a single node considered in this paper. While we assume surface codes on 2D devices~\cite{fowler2012surface,fowler2018low} and fault-tolerant communication with remote lattice surgery~\cite{horsman2012surface,ramette2024fault,fowler2010surface}, our proposal can be applied to more generic cases with quantum low-density parity-check codes~\cite{bravyi2024high,xu2024constant}, more efficient fault-tolerant communications~\cite{pattison2025constant,ataides2025constant,maeda2025logical}, and devices with more flexible connectivity such as shuttling~\cite{bluvstein2024logical,cai2023looped,siegel2025snakes}. See Sec.\,\ref{sec:discussion} for the compatibility with such theoretical proposals.

If we can assume a single-node FTQC~(The right regime of Fig.\,\ref{fig:overview}), the execution time is roughly estimated as $D T_{\text{local}}$, which we refer to as the algorithmic execution time. Our goal is to provide a DFTQC design that can make the total execution time close to the algorithmic execution time by leveraging a large node count $N$. 

It should be noted that one of the most studied regimes of distributed quantum computing is a regime where each node has a few \textit{physical} qubits. This situation is well studied as a popular regime of distributed QEC~\cite{campbell2007distributed,nickerson2013topological,fujii2012distributed,li2012high} or measurement-based quantum computation~\cite{bombin2021interleaving,bourassa2021blueprint,fukui2020temporal,larsen2021fault}~(The left regime of Fig.\,\ref{fig:overview}). These systems aim to implement a single block of a QEC code across multiple nodes. The requirement for quantum communication in this setting is entirely different, as fast and high-fidelity quantum communication is necessary to reduce logical error rates. In contrast, our DFTQC regime does not require such high-fidelity quantum communication because logical entanglement can be distilled. Although distributed QEC might be a viable candidate, it is outside the scope of this paper.

\subsection{Existing approaches}
Here, we review two primary directions for constructing DFTQCs: circuit-based DFTQC and measurement-based DFTQC.

\subsubsection{Circuit-based DFTQC}
The most naive implementation of DFTQC on small nodes is to assign each algorithmic qubit to a specific node. We refer to this approach as circuit-based DFTQC~(CB-DFTQC). The total execution time is roughly $D T_{\text{Bell}}$, which is much slower than the algorithmic execution time $D T_{\text{local}}$. In this paper, we refer to this time as a baseline of CB-DFTQC. 

To mitigate the execution time of CB-DFTQC, extensive efforts have been made to reduce the number of non-local logical operations.
When each algorithmic qubit is assigned to a specific node, the communication overhead can be reduced by optimizing the qubit-to-node assignment. This problem can often be formulated as a graph partitioning problem~\cite{andres2019automated,zomorodi2018optimizing}. Additional reductions in communication can be achieved by temporarily transferring qubit data between nodes during execution~\cite{wu2022autocomm}.
If the network topology can be designed according to the structure of the given quantum circuit, further optimizations become possible. In particular, if optical switches with acceptable insertion loss are available, efficient dynamically reconfigurable network architectures can be considered~\cite{zhang2024optimizing,main2025distributed}.
Alternatively, quantum circuits can be compiled from high-level descriptions into instruction sequences to provide better partitioning solutions~\cite{yoshioka2024hunting}. There are several other strategies to reduce the cost of partitioning using additional logical qubits~\cite{daei2020optimized,nikahd2021automated,burt2024generalised,andres2024distributing,sundaram2022distribution,sundaram2023distributing,cambiucci2023hypergraphic,baker2020time,ovide2023mapping,bandic2023mapping,pastor2024circuit}.

Despite massive efforts in this direction, most existing techniques assume a regime in which each node can store several tens of logical qubits. In addition, many of them are developed primarily for the noisy intermediate-scale quantum regime or abstract distributed quantum computing with a given network structure. Consequently, their effectiveness is limited in the regime where each node holds only a few logical qubits, i.e., when $n_{\text{node}} \sim 1$.

\subsubsection{Measurement-based DFTQC}
Another paradigm developed in the quantum computing community is the idea of measurement-based quantum computation~(MBQC)~\cite{raussendorf2001one,raussendorf2003measurement}. In this paradigm, we prepare a cluster state whose size depends on the algorithmic qubit count and depth, and perform computation via repeated quantum teleportation with measurements in an appropriate basis. By assigning each node of cluster state to a node of DFTQC, we can implement MBQC on many small-node DFTQCs, which we refer to as measurement-based DFTQC~(MB-DFTQC). 

\begin{figure}[t]
    \centering
    \begin{minipage}{0.9\linewidth}
        \centering
        \includegraphics[width=\linewidth]{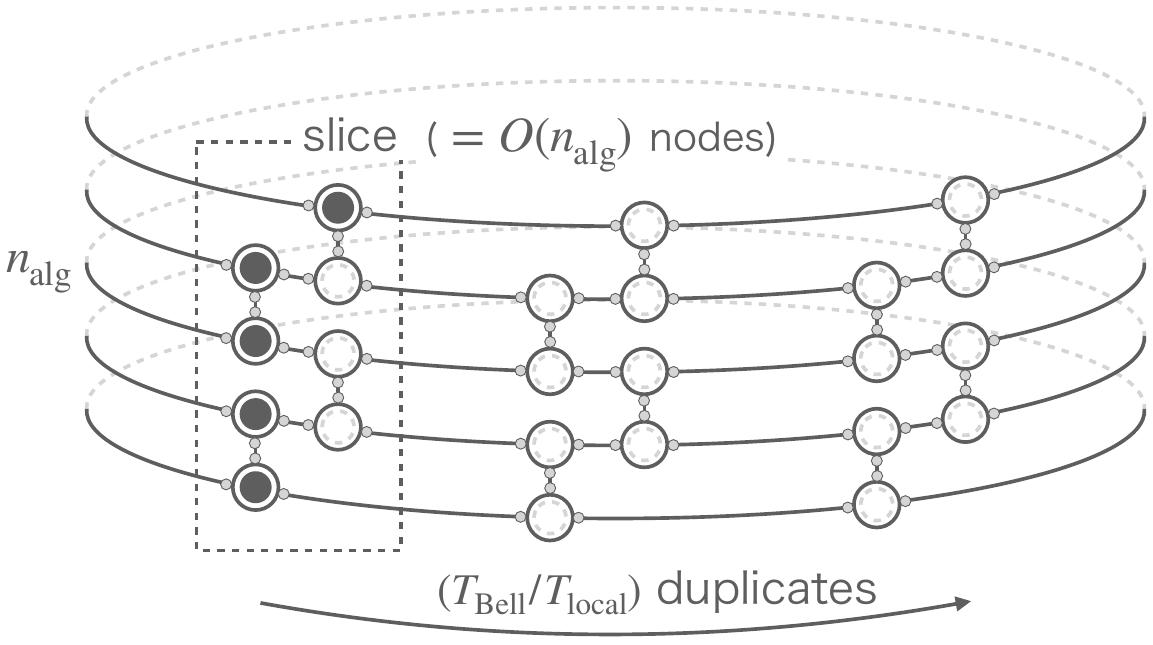}
        \subcaption{}
        \label{fig:brickwork_MB_DFTQC}
    \end{minipage}
    \begin{minipage}{0.9\linewidth}
        \centering
        \includegraphics[width=\linewidth]{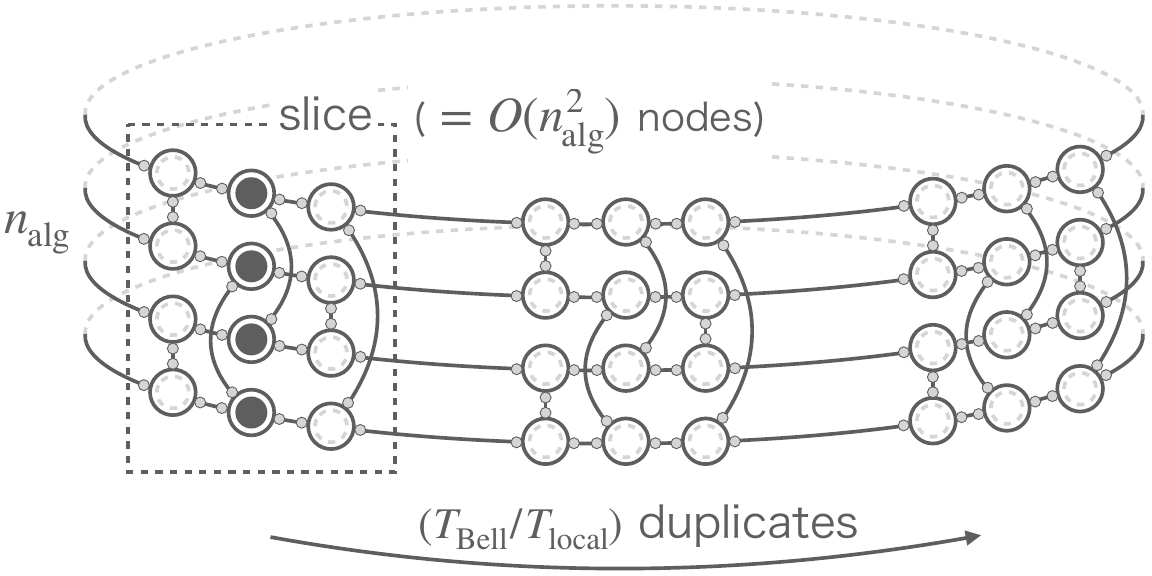}
        \subcaption{}
        \label{fig:clique_MB_DFTQC}
    \end{minipage}
    \caption{MB-DFTQC implementations with ring-shaped cluster states. (a) With the brickwork cluster state, the execution time scales as $O(T_{\text{Bell}} + n_{\text{alg}} D T_{\text{local}})$ and the node count scales as $O(n_{\text{alg}} T_{\text{Bell}} / T_{\text{local}})$. (b) If we use the clique cluster state, the execution time is improved to $O(T_{\text{Bell}} + D T_{\text{local}})$ but the node count scales as $O(n_{\text{alg}}^2 T_{\text{Bell}} / T_{\text{local}})$.}
    \label{fig:MB_DFTQC}
\end{figure}

A significant advantage of MBQC is that its execution time is almost independent of the generation time of logical Bell pairs. Once a cluster state is generated at the beginning of MBQC, the remaining execution time depends only on local logical operations and classical communications.
There are two major concerns in MB-DFTQC. One is that it is ambiguous how to manage probabilistic resource generation, such as magic-state preparation. The other is the significant increase in the number of required nodes. The latter is particularly serious as the number of nodes in the cluster states in MBQC is at least proportional to the area of quantum circuits $O(n_{\text{alg}} D)$. If we can reuse the node after measurement, the node count can be reduced by continuously generating and consuming entangled nodes in ring-shaped networks~\cite{monroe2014large}. While Ref.\,\cite{monroe2014large} assumes each node of MBQC contains a few physical trapped-ion qubits, this idea can be straightforwardly extended to the case where each node contains a few logical qubits rather than a few physical qubits, as shown in Fig.\,\ref{fig:brickwork_MB_DFTQC}.
The ring of cluster states must be sufficiently long so that the time for generating Bell pairs is shorter than consuming a round of cluster states. Thus, the length of the ring can be estimated as $T_{\text{Bell}} / T_{\text{local}}$.

It is known that several types of cluster states, such as the brickwork cluster state~\cite{broadbent2009universal} shown in Fig.\,\ref{fig:brickwork_MB_DFTQC}, can perform universal quantum computation. Since the brickwork cluster state assumes that each node stores a single algorithmic qubit and has quantum channels to three other nodes, we can straightforwardly implement MB-DFTQC with many small fault-tolerant nodes even with $n_{\text{node}} = 1$ and $d=3$. Since MBQC with brickwork cluster states can be translated into circuit-based quantum computing with linear connectivity, it requires $O(n_{\text{alg}})$ operations to perform each non-local two-qubit gate on average. Thus, the time complexity scales as $O(T_{\text{Bell}} + n_{\text{alg}} D T_{\text{local}})$ and the node count scales as $O(n_{\text{alg}} T_{\text{Bell}} / T_{\text{local}})$. 
This execution-time penalty can be mitigated by using densely coupled cluster states. Figure\,\ref{fig:clique_MB_DFTQC} shows the clique cluster state as an example, where the all-to-all connectivity among algorithmic qubits is guaranteed in each slice. In this case, each slice uses $O(n_{\text{alg}}^2)$ nodes, and we need $T_{\text{Bell}} / T_{\text{local}}$ slices for concealing the Bell generation latency. Thus, the node scaling becomes $O(n_{\text{alg}}^2 T_{\text{Bell}} / T_{\text{local}})$ while achieving nearly algorithmic execution time $O(T_{\text{Bell}} + D T_{\text{local}})$. This node count remains challenging to demonstrate; however, it is non-trivial to further optimize the number of nodes under the MBQC framework.

\subsubsection{Summary and Motivation}
The existing consideration of DFTQC can be categorized into two paradigms. 
The first paradigm, circuit-based DFTQC (CB-DFTQC), is based on embedding quantum circuits into a node network. This approach exhibits good node-count scaling, and we can introduce optimizations inspired by those developed in classical distributed computing. On the other hand, the execution time is significantly increased when we use small nodes with narrow bandwidth.

The second paradigm, measurement-based DFTQC (MB-DFTQC), is built upon the measurement-based quantum computing (MBQC) framework. In this paradigm, quantum computation is divided into a program-independent cluster-state generation phase and a program-dependent state-teleportation phase. Because this framework leverages the intrinsic properties of quantum teleportation, it has been extensively studied as a distinct form of quantum computation. While this paradigm makes the total execution time almost independent of communication bandwidth, it still incurs overheads in node count when simulating quantum circuits with all-to-all connectivity. These overheads arise from the constraint that the cluster state must be independent of the gate sequence. This assumption is often adopted because MBQC is typically considered in the context of optical platforms~\cite{bombin2021interleaving,bourassa2021blueprint,fukui2020temporal,larsen2021fault}, where it is difficult to reuse the measured qubits. On the other hand, this restriction is unnecessary in the design of DFTQCs.

In summary, CB-DFTQC is straightforward but imposes significant communication overhead in a many-small-node regime. Although the MB-DFTQC paradigm appears promising for minimizing communication overheads, it demands a huge number of fault-tolerant nodes. A strategy capable of realizing fast DFTQC with small FTQC nodes connected by slow logical quantum links has yet to be established. Therefore, it is highly desirable to develop a high-performance architecture tailored to this regime.

\begin{table*}[t]
    \centering
    \begin{tabular}{l|l|l}
        \hline
        Type & Time scaling & Node scaling \\
        \hline
        \hline
        Algorithmic circuit & $O\left(D T_{\text{local}}\right)$ & - \\
        \hline
        \hline
        CB-DFTQC
        & $O\left(D T_{\text{Bell}}\right)$
        & $O\left(n_{\rm alg}\right)$ \\
        \hline
        MB-DFTQC (brickwork)
        & $O\left(T_{\text{Bell}}+ n_{\rm alg} D T_{\text{local}}\right)$ 
        & $O\left(n_{\rm alg} D\right)$ \\
        \hline
        MB-DFTQC (brickwork, ring)
        & $O\left(T_{\text{Bell}}+ n_{\rm alg} D T_{\text{local}}\right)$ 
        & $O\left(n_{\rm alg} (T_{\text{Bell}} / T_{\text{local}})\right)$ \\
        \hline
        MB-DFTQC (clique)
        & $O\left(T_{\text{Bell}}+ D T_{\text{local}}\right)$ 
        & $O\left(n_{\rm alg}^2 D\right)$ \\
        \hline
        MB-DFTQC (clique, ring)
        & $O\left(T_{\text{Bell}}+ D T_{\text{local}}\right)$ 
        & $O\left(n_{\rm alg}^2 (T_{\text{Bell}} / T_{\text{local}})\right)$ \\
        \hline
        \hline
        NBQC (circuit-agnostic) 
        & $O\left(T_{\text{Bell}}+ D T_{\text{local}}\right)$ 
        & $O\left(n_{\rm alg}^2 (T_{\text{Bell}}/T_{\text{local}})\right)$ \\
        \hline
        NBQC (general form) 
        & $O\left(T_{\text{Bell}}+ D T_{\text{local}}\right)$ 
        & $O\left(\sum_{i=1}^{n_{\rm alg}} (\sum_j \braket{\rm Bias}_{i,j})^{\log_s t}\right)$ \\
        \hline
        NBQC (uniform access) 
        & $O\left(T_{\text{Bell}}+ D T_{\text{local}}\right)$ 
        & $O\left(n_{\rm alg} (T_{\text{Bell}} / T_{\text{local}})^{\log_s t}\right)$ \\
        \hline
        NBQC (biased access) 
        & $O\left(T_{\text{Bell}}+ D T_{\text{local}}\right)$ 
        & $O\left((T_{\text{Bell}} / T_{\text{local}})^{\log_s t}\right)$ \\
        \hline
    \end{tabular}
    \caption{Comparison between our proposal and the existing frameworks in terms of time and node-count scaling. Here, $n_{\text{alg}}$ is the number of algorithmic qubits, $D$ is the algorithmic depth, $T_{\text{local}}$ is the latency of local logical operations, and $T_{\text{Bell}}$ is the latency of logical entanglement generation. $\braket{\rm Bias}_{i,j}$ represents the maximum number of communications between the $i$-th and the $j$-th components occurring in $T_{\text{Bell}}$ of time duration (see Sec.\,\ref{sec:circuit-specific-bottleneck-free-design} for the definition). The parameter $(s,t)$ is a small constant that characterizes the Clos network (explained later). We typically choose $(s,t)=(2,3)$ and an exponent is $\log_s t \sim 1.58$. Note that the spacetime complexity of magic state generation is omitted in this table for simplicity.}
    \label{tab:time_complexity}
\end{table*}

\section{Network-Based Quantum Computation}
\label{sec:nbqc}
\subsection{Overview}
\begin{figure*}[t]
    \centering
    \includegraphics[width=1.0\linewidth]{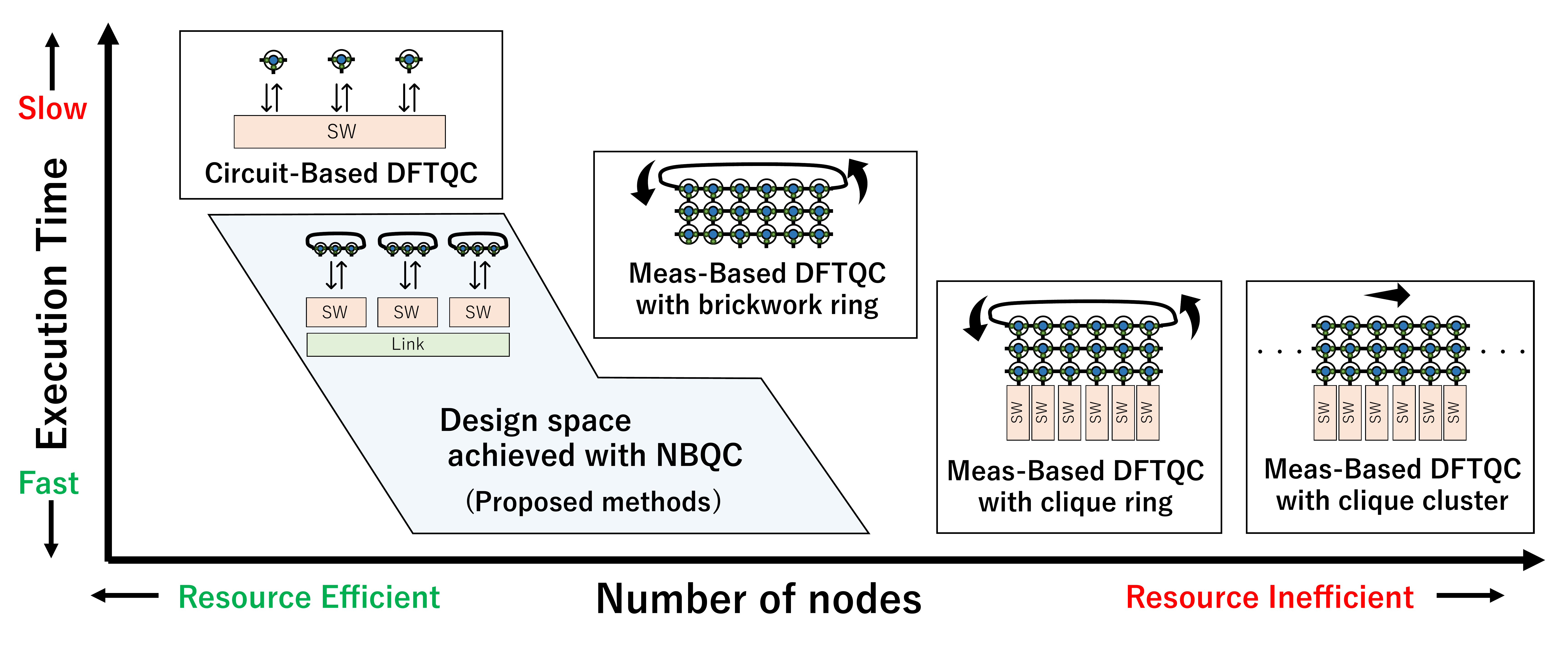}
    \caption{The achievable design space with CB-DFTQC, MB-DFTQC, and NBQC. The proposed method, NBQC, provides fast and resource-efficient designs compared to the existing approaches, and offers the tunability between execution time and node count.}
    \label{fig:positioning}
\end{figure*}

In this paper, we propose a novel design of DFTQC, named \textit{Network-Based Quantum Computation~(NBQC)}, for a regime where each node has $n_{\rm{node}} \sim 1$ algorithmic qubits and $d \sim 3$ slow quantum channels to other nodes. 
Table~\ref{tab:time_complexity} shows the scaling of execution time and node count of these designs compared to the existing approaches. Figure.\,\ref{fig:positioning} illustrates the design space achievable with the idea of NBQC. The qualitative comparison between CB-DFTQC, MB-DFTQC, and NBQC will be revisited in Sec.\,\ref{sec:discussion}.

We will present two types of DFTQC designs based on the idea of NBQC. One is \textit{circuit-agnostic NBQC}, which shows performance similar to MB-DFTQC with the clique-ring cluster state but explicitly accommodates probabilistic resource generations, such as magic-state generations. This means that our NBQC framework effectively contains MB-DFTQC as a special case. This design can show an algorithmic execution time for any quantum circuit.

The other design, \textit{circuit-specific NBQC}, is a further optimized design that is tailored to several properties of target applications, such as the access-frequency profile explained later. Compared to the existing approaches, this design can achieve a better trade-off between node count and execution time. In particular, if a small number of nodes are frequently accessed, the node count can be significantly reduced, as shown in Table~\ref{tab:time_complexity}. Also, circuit-specific NBQC can achieve a continuous trade-off between node count and time complexity. This enables us to design an efficient network for a given number of available nodes.

\begin{figure*}[t]
    \centering
    \includegraphics[width=0.9\linewidth]{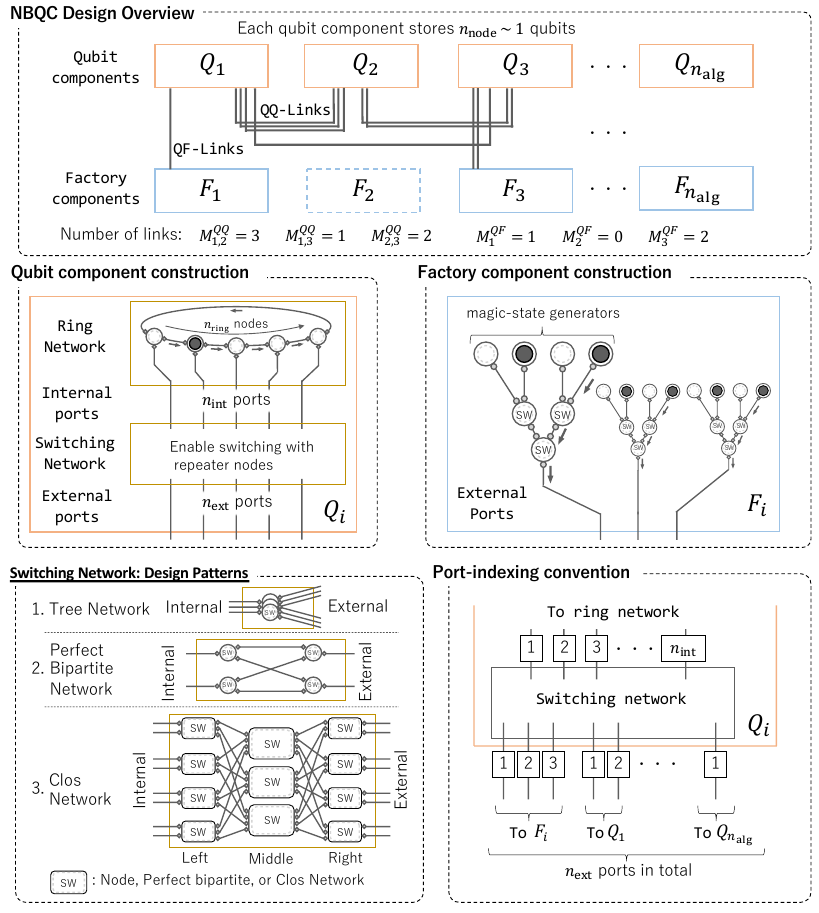}
    \caption{The whole design of the NBQC system.}
    \label{fig:NBQC_overview}
\end{figure*}

A key idea of NBQC is to partially introduce the concept of MB-DFTQC into the CB-DFTQC framework, aiming for shorter execution times with redundant nodes and flexible entanglement consumption. 
In CB-DFTQC, each algorithmic qubit remains on a specific node for most of the time. In this case, the rate of remote logical operations is limited by the speed of Bell pair generation.
To avoid this limitation, it is possible to allow each algorithmic qubit to move after consuming a Bell pair to provide it with another one.
MB-DFTQCs implicitly utilize this idea to remove the latency for entanglement generation from execution time effectively. However, in the case of MB-DFTQC, it is difficult to reduce the number of nodes according to the structures of circuits since all the algorithmic qubits must teleport synchronously along with a single ring-shaped cluster state~(Fig.\,\ref{fig:MB_DFTQC}).
In NBQC, as illustrated in Fig.~\ref{fig:NBQC_overview}, we assign a ring network to each algorithmic qubit and modify its length independently according to the communication frequency. If the length of the ring is sufficiently long, the node with the algorithmic qubit always has a distilled logical Bell pair. To support any pattern of remote two-qubit gates between algorithmic qubits, we attach a strict-sense non-blocking switching network~(explained in Sec.\,\ref{sec:nbqc}) to each ring network, which guarantees the existence of a chain of Bell pairs between any pair of algorithmic qubits in two different ring networks.

NBQC materializes this idea with two types of components: the qubit component and the factory component, as shown at the top of Fig.\,\ref{fig:NBQC_overview}. The qubit component contains a single code block for algorithmic qubits (i.e., it contains a single algorithmic qubit if $n_{\rm node}=1$), and the factory component generates magic states. These components are connected by links between two qubit components (QQ-link) and by links between qubit and factory components (QF-link), and the ports of components connected by links are called {\it external ports}. The qubit component contains a ring network that enables continuous quantum-state teleportation, ensuring that algorithmic qubits are always placed at a node with distilled logical entanglement. Since the position of the algorithmic qubit in the ring network depends on the time, a switching network is inserted before the external ports to allow flexible access. We refer to ports between the ring network and the switching network \textit{internal ports}. The factory component consists of several trees of nodes, and each leaf node is called a \textit{magic-state generator}, which repeats the probabilistic generation of magic states. 

\subsection{Qubit Component}
Each qubit component stores one or several algorithmic qubits. In the following explanation, we again note that we focus on a case with $n_{\rm node}=1$ for simplicity, i.e., each qubit component stores a single algorithmic qubit. This can be straightforwardly extended to cases $n_{\rm node}>1$ as explained in Sec.\,\ref{sec:discussion}. This component facilitates high-bandwidth communication with the other qubit components and a factory component.
A qubit component consists of a {\it ring network} as a storage, and a {\it switching network} as a router, both of which are introduced in the following sections.

\subsubsection{Ring network}
The ring network consists of a 1D ring of nodes as shown in Fig.\,\ref{fig:overview}. Two channels per node are connected to form a ring, and the remaining channels are connected to the switching network through internal ports. The algorithmic qubit resides on a specific node in the ring network. At the beginning of NBQC, the algorithmic qubit is in the $0$-th node. After waiting for $T_{\text{Bell}}$, each channel has a single distilled logical Bell pair. 

Local Clifford operations (e.g., logical $H$ and $S$ gates) are executed inside the node. Remote operations (e.g., a CNOT gate, a two-qubit Pauli measurement, or gate teleportation consuming magic states) are executed by consuming a chain of logical entanglements connecting the algorithmic qubit in the ring network to another node outside of the qubit component via the switching network.
If all the Bell pairs between the residing node and the switching network are consumed, the algorithmic qubit is moved from the $i$-th node to $(i+1 \mod n_{\text{ring}})$-th node by quantum teleportation, where $n_{\text{ring}}$ is the length of the ring network.

After the teleportation is repeated $n_{\text{ring}}$ times, the algorithmic qubit returns to the $0$-th node. Since remote operations to other components and quantum state teleportation to the neighboring ring-network node take $T_{\text{local}}$ for consuming a Bell pair, the algorithmic qubit stays at each node at least $(d-1)T_{\text{local}}$, and a round-trip of the ring network takes at least $n_{\text{ring}}(d-1)T_{\text{local}}$. Therefore, by choosing $n_{\text{ring}}=\lceil T_{\text{Bell}} / ((d-1)T_{\text{local}}) \rceil$, the round-trip time becomes longer than the latency of entanglement generation $T_{\text{Bell}}$, and the node with the algorithmic qubit can always consume a distilled Bell pair to the switching network without waiting for the entanglement regeneration.

It should be emphasized that it is not always necessary to choose a large $n_{\text{ring}}$ for all the qubit components; if the $i$-th algorithmic qubit rarely becomes a target of remote operations, there is no execution time overhead in choosing small $n_{\text{ring}}$ for the $i$-th qubit component.

\subsubsection{Switching network}
When remote logical operations or non-Clifford gates are executed, a communication path must exist between the two nodes that hold the corresponding algorithmic qubits or magic states. The \textit{switching network} connects the ring network to other components.

For each remote logical operation, we must assign an entanglement-swapping path using channels that already possess distilled logical Bell pairs. For any two remote operations executed within a time window $T_{\mathrm{Bell}}$, their entanglement-swapping paths must be edge-disjoint. Otherwise, a subsequent remote operation must wait for the regeneration of logical Bell pairs, introducing additional latency. Thus, the aim of switching network design is to prevent such latency while minimizing the number of required repeater nodes.

We can formulate this requirement as a graph problem. Let the number of internal and external ports be denoted by $n_{\rm int} = n_{\mathrm{ring}}(d-2)$ and $n_{\rm ext}$, respectively. We consider a bipartite graph in which the internal ports are represented by $V_I = (1, \ldots, n_{\rm int})$ and the external ports by $V_E = (1, \ldots, n_{\rm ext})$. A set of pairs $M \subseteq V_I \times V_E$ is called a matching if the pairs are vertex-disjoint; that is, for any two pairs $(v_I, v_E), (v_I', v_E') \in M$, we have $v_I \neq v_I'$ and $v_E \neq v_E'$.

\begin{table}[t]
\begin{tabular}{ccc}
\hline
Type                      & Connectivity                                                        & Number of nodes                       \\ \hline
Tree network              & limited                                                             & $O(n_{\rm ext})$                      \\
Perfect bipartite network & \begin{tabular}[c]{@{}c@{}}strict-sense\\ non-blocking\end{tabular} & \textbf{$O(n_{\rm int} n_{\rm ext})$} \\
Clos network              & \begin{tabular}[c]{@{}c@{}}strict-sense\\ non-blocking\end{tabular} & $O(sn_{\rm int}^{\log_s(2s-1)})$                \\ \hline
\end{tabular}
\caption{Three types of switching networks. Note that we assume each external port is connected to exactly one internal port in the ``Tree network'' row. Also, we assume $n_{\rm int} = n_{\rm ext}$ at the ``Clos network'' row for simplicity.}
\label{tab:switching_network_comparison}
\end{table}

Here, we introduce three types of networks: tree network, perfect bipartite network, and Clos network, as shown Table~\ref{tab:switching_network_comparison} and the left bottom of Fig.\,\ref{fig:NBQC_overview}.
The first type, tree network, supports a case where the $i$-th internal port is connected to a subset of external ports $V_E^{(i)} \subseteq V_E$, and external ports $V_E^{(i)}$ are connected only to the $i$-th port. We define $N_{\rm tree}(n; d)$ as the number of nodes required for a tree network with one root port and $n$ leaf ports. Then, $N_{\rm tree}(n; d)$ is given as
\begin{align}
    N_{\rm tree}(n; d) = \begin{cases}
        1 & \text{if } n < d \\
        \ceil{\frac{n-1}{d-2}} & \text{if } n \ge d.
    \end{cases}
\end{align}
The node count for the tree network connecting $n_{\rm int}$ internal ports and $n_{\rm ext}$ external ports, denoted as $N_{\rm tree}(n_{\rm int}, n_{\rm ext}; d)$ can be calculated as
\begin{align}
    N_{\rm tree}(n_{\rm int}, n_{\rm ext}; d) = \sum_{i=1}^{n_{\rm int}} N_{\rm tree}(|V_E^{(i)}|; d),
\end{align}
where $|V_E^{(i)}|$ represents the size of $V_E^{(i)}$.

The other two types, the perfect bipartite network and the Clos network, are used for cases where all internal ports may be connected to all external ports. In such cases, the switching network must have the \textit{strict-sense non-blocking} property. This property is guaranteed if the following conditions are satisfied:
\begin{itemize}
\item Condition~1: For any given matching $M$, there is a set of edge-disjoint paths from $v_I$ to $v_E$ for every $(v_I, v_E) \in M$. In other words, we can assign paths to all the given pairs of ports on both sides, and we can perform entanglement swapping from the ring network to the other component by consuming independent Bell pairs.
\item Condition~2: Suppose that elements of a matching $M$ are revealed one by one, and we need to assign a path to each element without knowing the subsequent elements. Then, there is an algorithm to provide a path to each element satisfying Condition~1.
\end{itemize}

\begin{figure}[t]
    \begin{minipage}{0.9\linewidth}
        \centering
        \includegraphics[width=\linewidth]{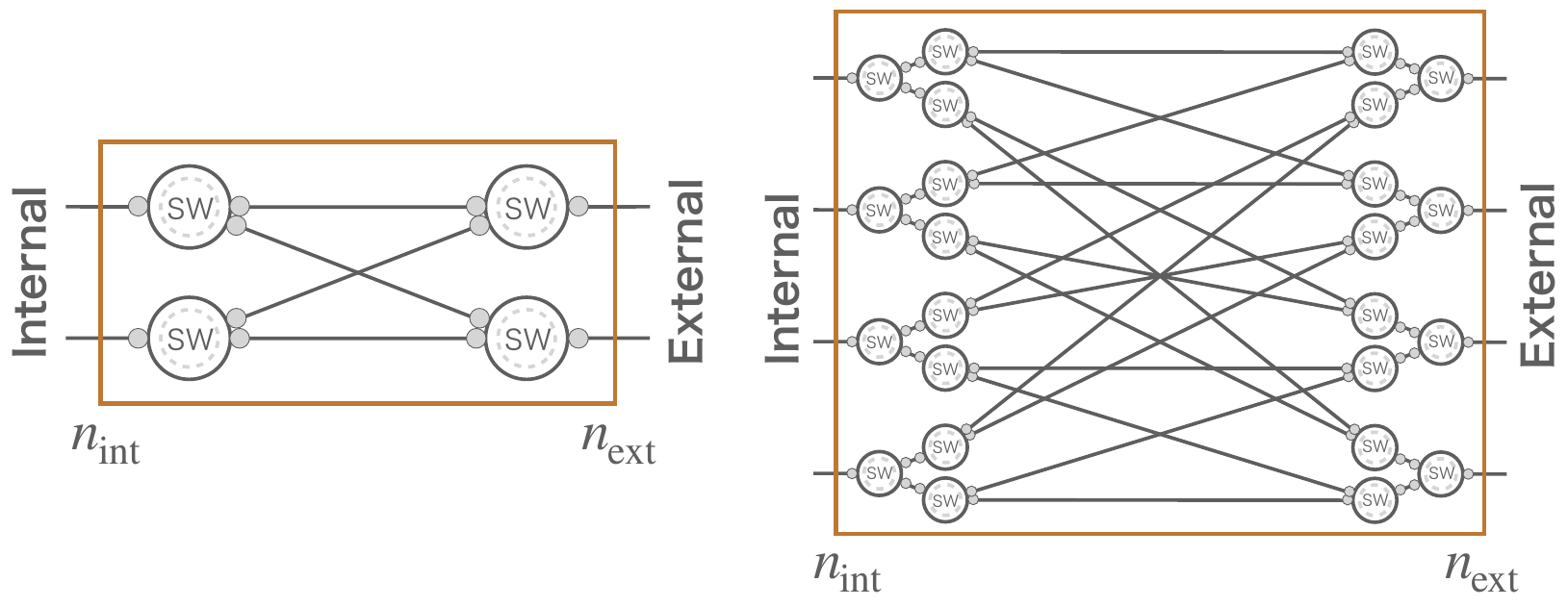}
        \subcaption{}
        \label{fig:bipartite_network}
    \end{minipage}
    \begin{minipage}{0.9\linewidth}
        \centering
        \includegraphics[width=\linewidth]{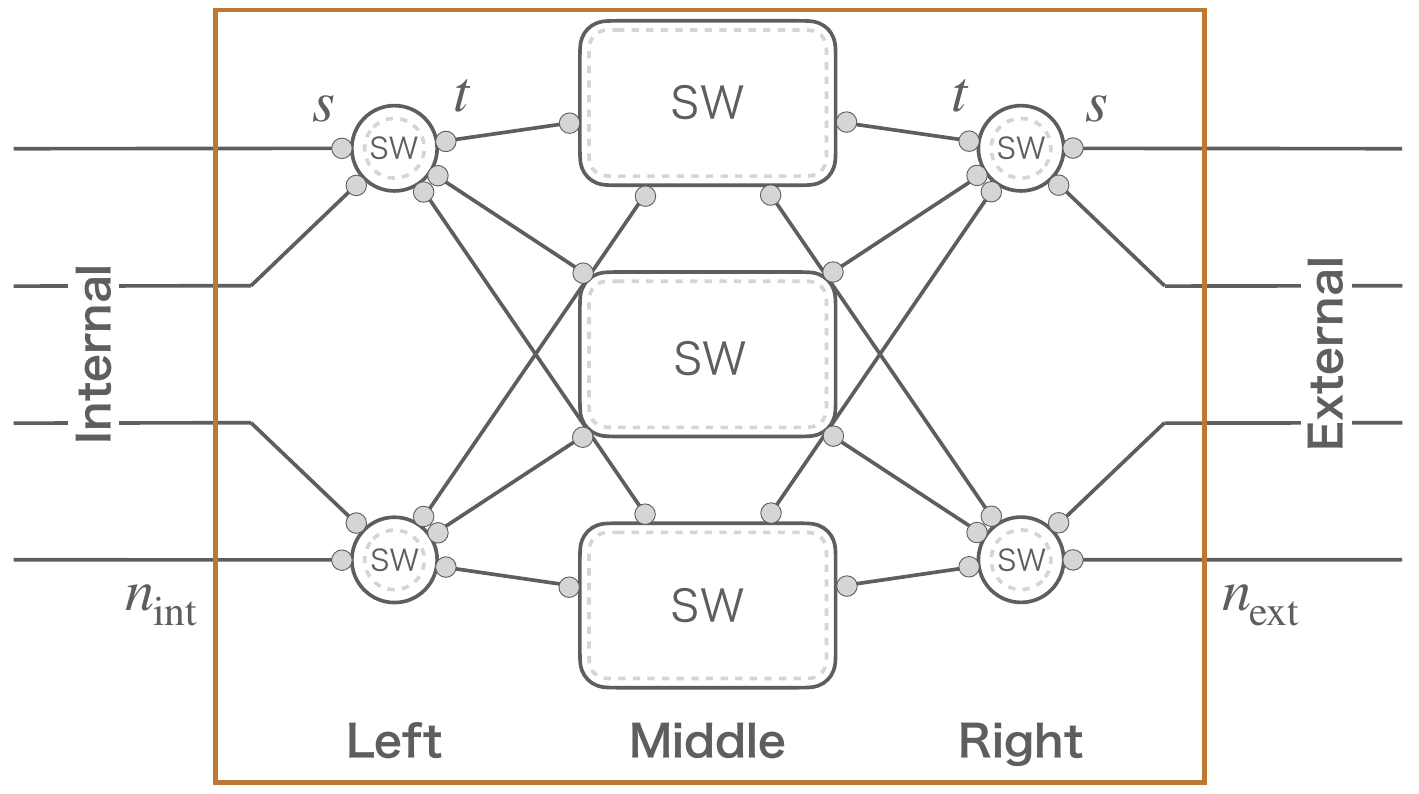}
        \subcaption{}
        \label{fig:clos_network}
    \end{minipage}
    \caption{(a)Perfect bipartite network.~\textbf{left:} When the required degree of each node is less than $d$, like the case of $n_{\rm int} = n_{\rm ext} = 2$ and $d=3$, we can construct a perfect bipartite graph with $n_{\rm int}$ and $n_{\rm ext}$ nodes on each side.~\textbf{right:} When the required degree of each node is more than $d$, like the case of $n_{\rm int} = n_{\rm ext} = 4$ and $d=3$, we can replace the node with a tree.
    (b)Clos network. The design parameter used in this example is $(s, t) = (2, 3)$. Note that each switch can be replaced with a perfect bipartite network when $d$ is small.}
    \label{fig:switching_networks}
\end{figure}

A straightforward way to construct a strict-sense non-blocking network is to make the perfect bipartite network with $n_{\rm int}$ and $n_{\rm ext}$ ports on each side. Since it can contain any bipartite network as its subgraph, its strict-sense non-blocking property is guaranteed. The perfect bipartite network can be constructed with a single node if $d \geq n_{\rm int} + n_{\rm ext}$. In case where $n_{\rm int} + n_{\rm ext} > d > {\rm max}(n_{\rm int}, n_{\rm ext})$, we assign a node for each internal and external port. Then, each node on the internal-port side can be connected to every node on the external-port side, as shown on the left of Fig.\,\ref{fig:bipartite_network}. If ${\rm max}(n_{\rm int}, n_{\rm ext})\ge d$, for each internal port, we put a tree network $N_{\rm tree}(n_{\rm ext}; d)$ and set its root port as the internal port. We also assign $N_{\rm tree}(n_{\rm int}; d)$ to each external port similarly. Then, leaf ports of the tree network connected to an internal port are connected to a leaf port connected to every external port. An example is shown in the right of Fig.\,\ref{fig:bipartite_network}. The number of nodes required for the perfect bipartite network, denoted as $N_{\rm bipartite}(n_{\rm int}, n_{\rm ext}; d)$, can be calculated as
\begin{align}
    & N_{\rm bipartite}(n_{\rm int}, n_{\rm ext}; d) \nonumber\\
    & = \begin{cases}
        1 & \text{if } n_{\rm int} + n_{\rm ext} \leq d \\
        n_{\rm ext} N_{\rm tree}(n_{\rm int}; d) \nonumber \\
        + n_{\rm int} N_{\rm tree}(n_{\rm ext}; d) & \text{if } n_{\rm int} + n_{\rm ext} > d,
    \end{cases}
\end{align}
which scales as $O(n_{\rm int} n_{\rm ext})$ when $n_{\rm int}$ and $n_{\rm ext}$ are sufficiently larger than $d$.

A more efficient approach to constructing a strict-sense non-blocking network is to employ a Clos network~\cite{clos_network}. As illustrated in Fig.\,\ref{fig:clos_network}, the Clos network comprises three stages, left, middle, and right stages, each consisting of an array of switches, and each switch is composed of nodes. 
A switch in the left stage has $s \geq 2$ ports on the ring-node side and $t$ ports on the opposite side, where $s$ and $t$ are design parameters. If $s + t \le d$, each switch corresponds to a single node. Otherwise, each switch is constructed as a perfect bipartite network with $N_{\rm bipartite}(s, t; d)$ nodes. Since there are $n_{\rm int}$ ports from the ring network, the left stage contains $\ceil{n_{\rm int} / s}$ switches. The right stage is similarly defined and consists of $\ceil{n_{\rm ext} / s}$ switches. 

The middle stage contains $t$ switches, each having $\ceil{n_{\rm int} / s}$ ports connected to the left stage and $\ceil{n_{\rm ext} / s}$ ports connected to the right stage. Every switch in the left and right stages is fully connected to all the switches in the middle stage. If $\ceil{n_{\rm int} / s} + \ceil{n_{\rm ext} / s} > d$, the switches in the middle stage are recursively defined using smaller Clos networks or perfect bipartite networks.
It is known that the strict-sense non-blocking property is guaranteed when $t \ge 2s - 1$~\cite{clos_network}. 

The upper bound of the node count in the Clos network can be evaluated as follows. We define $k$ as the minimum integer satisfying $\max\left( n_{\rm int}, n_{\rm ext} \right) \leq s^k$. Then, the node-count scaling of the Clos network $N_{\rm Clos}(n_{\rm int}, n_{\rm ext}; d)$ can be upper-bounded by $N_{\rm Clos}(s^k, s^k; d)$. The number of switches in the left and the right stage is $s^{k-1}$, and the middle stage consists of $t$ switches connecting $s^{k-1}$ internal and external ports. Thus, $N_{\rm Clos}(s^k, s^k; d) = 2s^{k-1} N_{\rm bipartite}(s, t; d) + t N_{\rm Clos}(s^{k-1}, s^{k-1}; d)$ holds for $k > 1$. Assuming that we use a perfect bipartite network when $k=1$, i.e., $N_{\rm Clos}(s, s; d) = N_{\rm bipartite}(s, s; d)$, we can estimate the total number of nodes as $N_{\rm Clos}(s^k, s^k; d) = \frac{2s}{t-s}(t^{k-1}-s^{k-1})N_{\rm bipartite}(s, t; d) + t^{k-1}N_{\rm bipartite}(s, s; d)$. As we can assume $t > s$ for strict-sense non-blocking Clos networks, the order can be evaluated as $O(s t^k)$. If we choose $(s,t)$ as a small constant, by using $t^k = (s^k)^{\log_s t}$ and $s^k < s \max(n_{\rm int}, n_{\rm ext})$, we can find an upper bound $N_{\rm Clos}(n_{\rm int}, n_{\rm ext}; d) = O(\max(n_{\rm int}, n_{\rm ext})^{\log_s t})$. If we choose the minimum constant $(s,t)=(2,3)$ for example, the exponential factor is $\log_2 3 \simeq 1.58$.

\subsection{Factory Component}
\begin{figure}[t]
    \centering
    \includegraphics[width=0.8\linewidth]{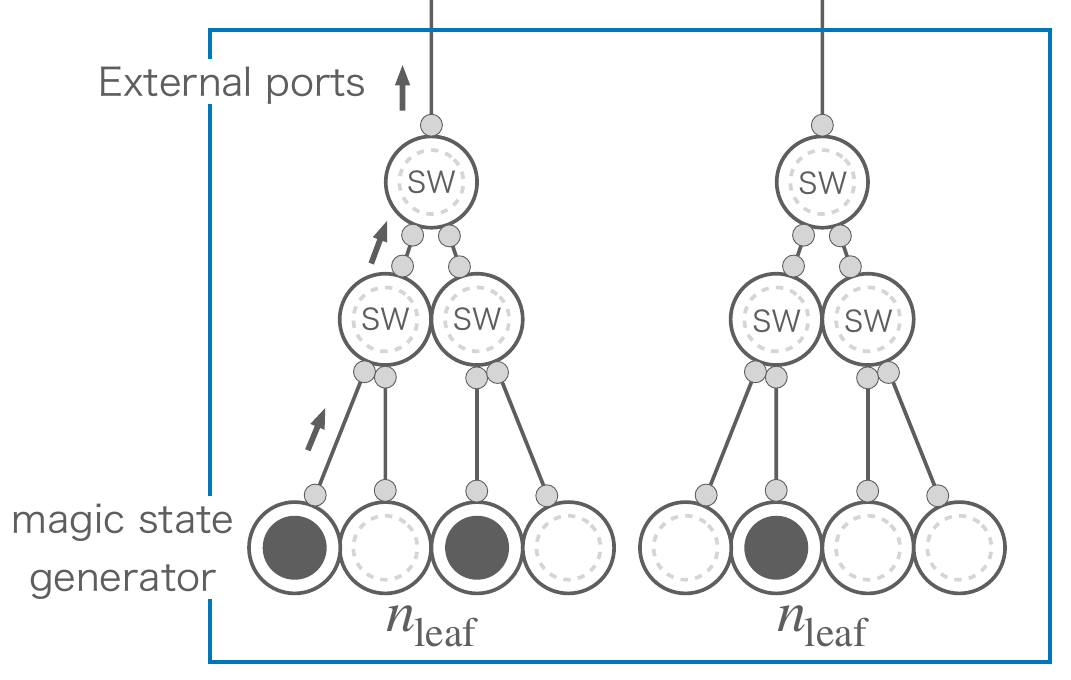}
    \caption{The construction of the factory component. Magic states are generated at the leaf nodes and transferred to the root node through the tree network. This example shows the case of $n_{\rm leaf} = 4$.}
    \label{fig:factory_component}
\end{figure}

The factory component is designed to generate magic states of $T$ gates and distribute them to a qubit component. As shown in Fig.\,\ref{fig:factory_component}, the factory component consists of {\it magic state generators} and {\it tree network}, both of which are constructed with nodes.
In NBQC, the $i$-th factory component is connected to the $i$-th qubit component and may have multiple external ports. 

We assume that each generator prepares magic states with a magic state cultivation protocol~\footnote{Note that if the fidelity of magic-state cultivation is not enough for computation, we need to perform magic-state distillation after the cultivation. In the regime considered in this paper, this distillation protocol must be performed as a sequence of non-local logical operations. Thus, we assume they are included in algorithmic quantum circuits.}. Once the cultivation protocol is successful, the magic state is stored inside the generator and is ready to be distributed.
The tree network connects its root to an external port and connects its leaves to magic state generators. When a magic state is requested through an external port, the factory component selects a generator that contains a magic state and finds a path between the external port and the generator. After that, the magic state is transferred via quantum state teleportation or directly consumed with remote logical operations.

We denote the time for a trial of magic-state cultivation and its success probability as $T_{\text{magic}}$ and $p_{\text{magic}}$, respectively. Let the number of generators per external port be $n_{\text{leaf}}$. Then, the trade-off between the number of nodes and the generation period of magic states per external port is given as follows.
In each time duration $T_{\text magic}$, $n_{\text{leaf}} p_{\text{magic}}$ magic states are successfully generated on average, and the average period for generating a single magic state is $T_{\text{magic}} / (n_{\text{leaf}} p_{\text{magic}})$. The supply rate is also upper-bounded by the generation rate of a logical Bell pair between the external port. Thus, the average period of providing magic states through an external port is $\max(T_{\text{magic}} / (n_{\text{leaf}} p_{\text{magic}}), T_{\text{Bell}})$. 

Therefore, choosing $n_{\text{leaf}} = \ceil{T_{\rm magic} / (T_{\text{Bell}} p_{\rm magic})}$ achieves the shortest period $T_{\text{Bell}}$, and choosing $n_{\text{leaf}}=1$ achieves the period $\max(T_{\rm magic}/p_{\rm magic}, T_{\text{Bell}})$. We can choose a desirable supply rate between them. If the supply period $T_{\text{Bell}}$ is insufficient, we can add external ports and tree networks to supply more magic states to the connected qubit component.

\subsection{Inter-Component Communication Links}
\begin{figure}[t]
    \centering
    \includegraphics[width=0.7\linewidth]{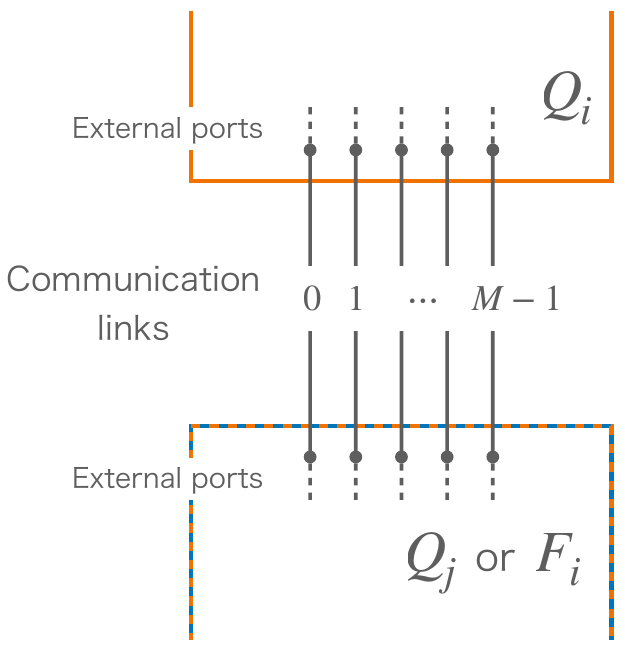}
    \caption{Inter-component communication links. The symbol ``$M$'' in this figure represents the number of links, which is equivalent to $M^{QQ}_{i,j}$ for QQ-links and $M^{QF}_{i}$ for QF-links.}
    \label{fig:communication_link}
\end{figure}

In NBQC, qubit components and factory components are connected via inter-component communication links. As illustrated in Fig.~\ref{fig:communication_link}, each communication link connects a pair of external ports on both components, and quantum communications are performed through it.

There are two types of communication links: \textit{QQ-link}, which connects two qubit components, and \textit{QF-link}, which connects a qubit component and a factory component.
We define $M^{QQ}_{i,j} \geq 0$ as the number of links between the $i$-th and $j$-th qubit components, and $M^{QF}_{i}$ as the number of links between the $i$-th qubit and factory components. 
Thus, the number of external ports of the $i$-th qubit component can be represented as $M^{QF}_{i} + \sum_{j \neq i} M^{QQ}_{i,j}$, and that of the $i$-th factory component as $M^{QF}_{i}$.

\subsection{Bottleneck-Free Design of NBQC}
In the above explanation, we introduced qubit components, factory components, and communication links connecting them. Since each component has several design flexibilities, there can be several designs of the whole NBQC. In this section, we show two types of NBQC designs, circuit-agnostic and circuit-specific designs. The former is a conservative design that guarantees the algorithmic execution time for any quantum circuit. The latter shows a better trade-off by assuming the access properties of remote operations in target applications.

With a sufficient number of redundant nodes, both designs can achieve execution times comparable to the algorithmic execution time. Concretely, we call an NBQC design that is free from the following execution-time overheads a {\it bottleneck-free design}, and both designs can be bottleneck-free with a sufficient number of nodes.
\begin{itemize}
    \item (Overhead 1) Inside a ring network, the algorithmic qubit is waiting for the generation of a Bell pair as it resides at a node without a distilled logical Bell pair.
    \item (Overhead 2) Inside a switching network, the regeneration of Bell pairs is required for finding a communication channel between internal and external ports.
    \item (Overhead 3) Inside a factory component, the preparation of magic states blocks the execution of $T$ gates.
    \item (Overhead 4) Between components, the generation of Bell pairs is required for inter-component communications.
\end{itemize}
In the following sections, we show that both designs can avoid these overheads with an appropriate choice of component designs.

\subsubsection{Circuit-Agnostic NBQC}
\label{sec:circuit-agnostic-bottleneck-free-design}
\begin{figure*}[t]
    \centering
    \includegraphics[width=0.7\linewidth]{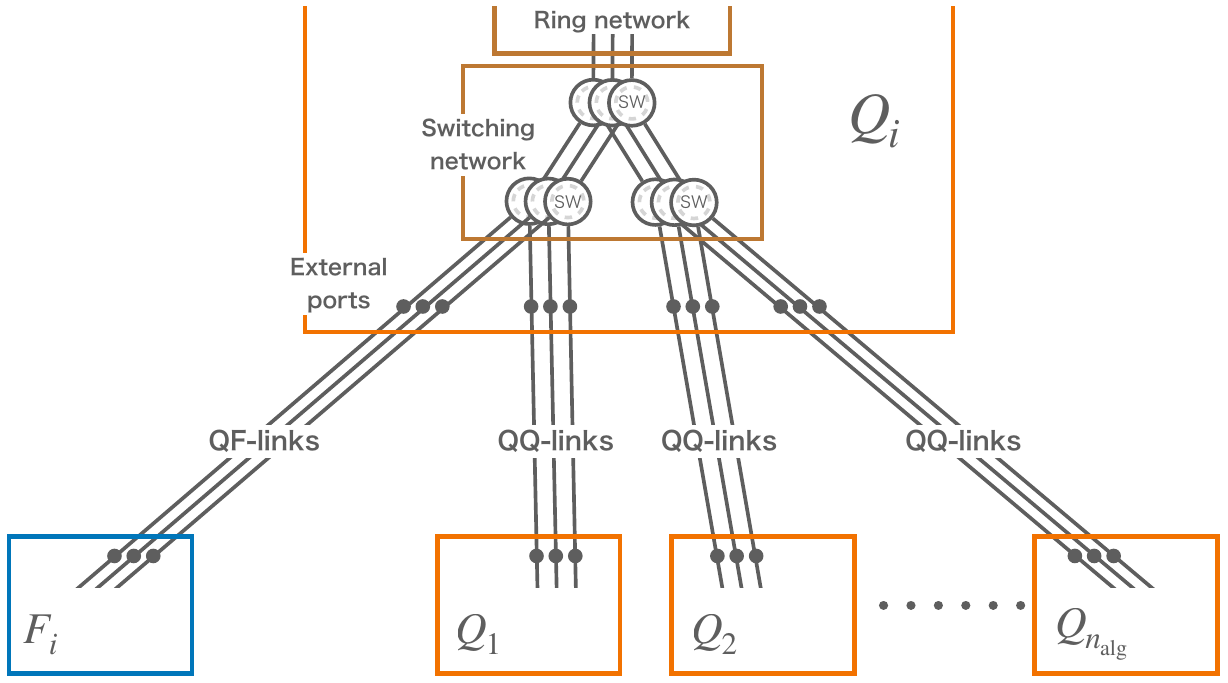}
    \caption{The qubit component in the circuit-agnostic NBQC. It is connected to $n_{\rm alg} - 1$ other qubit components and a factory component using QQ-links and QF-links. For each pair of components, $T_{\text{Bell}}/T_{\text{local}}$ QQ- or QF-links are used. Note that the switching network inside this qubit component does not need to have the strict-sense non-blocking property because we can assume that the positions of algorithmic qubits are synchronized.}
    \label{fig:circuit_agnostic_NBQC}
\end{figure*}

First, we show a bottleneck-free NBQC design without any assumptions on a target quantum circuit.
We refer to this design \textit{circuit-agnostic NBQC}. An example network structure is illustrated in Fig.\,\ref{fig:circuit_agnostic_NBQC}.
While the circuit-agnostic NBQC shows the same time and node-count scaling as MB-DFTQC, this design can explicitly handle probabilistic procedures, such as magic-state generations.

This design consists of $n_{\rm alg}$ qubit components and $n_{\rm alg}$ factory components. Every ring network in the qubit component has the common length $n_{\rm ring}=\lceil T_{\text{Bell}} / ((d-1)T_{\text{local}}) \rceil$. The number of internal ports becomes $n_{\rm int} = n_{\text{ring}}(d-2)$. Switching networks connect each internal port to $n_{\rm alg}$ independent external ports with a tree network, and thus $n_{\rm ext} = n_{\rm alg} n_{\rm int}$. The number of external ports of each factory component is the same as the number of internal ports of qubit components $n_{\rm int}$. Each external port of factory components is connected to a tree of $n_{\text{leaf}} = \ceil{T_{\rm magic} / (T_{\text{Bell}} p_{\rm magic})}$ magic-state generators.

These components are linked as follows. Inside the $i$-th qubit component, we label the $j$-th external port connected to the $k$-th internal port as $v_{i,j,k}$. Then, the external port $v_{i,i,k}$ is connected to the $k$-th external port of the $i$-th factory component. For $j \neq i$, $v_{i,j,k}$ is linked to $v_{j,i,k}$ via a QQ-link, i.e., the $i$-th external port connected to the $k$-th internal port inside the $j$-th qubit component. Thus, this design has $M_{i,j}^{QQ} = M_i^{QF} = n_{\rm int}$ QQ- and QF-links.

When we run the circuit-agnostic NBQC, we move the algorithmic qubit in each ring network synchronously, i.e., an algorithmic qubit in each qubit component resides at a node of the ring network with the same index. 
This construction can achieve a bottleneck-free design for the following reasons. As the length of each ring network is sufficiently long, each algorithmic qubit can move to the neighboring nodes without waiting for the generation of logical Bell pairs. Since every tree of the $k$-th internal port is used with an interval longer than $T_{\text{Bell}}$, we can assume that the switching network can always provide a path between qubit components. Since each factory component can also provide a magic state to each internal port in every $T_{\text{Bell}}$, the supply rate of magic states is never a bottleneck. Since the position of the algorithmic qubit is the same in each ring network, it is trivial to choose an inter-component communication link for given instructions.

If $d$ is a constant no less than three, we can see $n_{\rm int}=O(T_{\text{Bell}}/T_{\text{local}})$ and $n_{\rm ext} = O(n_{\rm alg} T_{\text{Bell}}/T_{\text{local}})$. Thus, each qubit and factory component consists of $O(n_{\rm alg} T_{\text{Bell}}/T_{\text{local}})$ nodes and $O(T_{\rm magic} / (T_{\text{local}} p_{\rm magic}))$ nodes, respectively.
Since there are $n_{\rm alg}$ qubit and factory components, the overall number of nodes for this design is $O\left(n_{\rm alg}^2 T_{\text{Bell}}/T_{\text{local}} + n_{\rm alg} T_{\rm magic} / (T_{\text{local}} p_{\rm magic})\right)$, which is the same complexity to that of the MB-DFTQC with clique-ring cluster states ignoring the cost for magic-state generators as listed in Table.\,\ref{tab:time_complexity}.

\subsubsection{Circuit-Specific NBQC}
\label{sec:circuit-specific-bottleneck-free-design}
Next, we discuss the construction of a bottleneck-free NBQC system based on the expected access-frequency profile of quantum circuits. The access-frequency profile is a list of the peak remote-operation frequency relative to the communication speed, which will be defined later.
Intuitively speaking, if a certain qubit rarely becomes a target of remote operations, reducing the nodes of the qubit component does not affect the execution time. Note that one may be concerned about the versatility of NBQC networks if they are tailored to a specific access-frequency profile. We expect this specification does not lose its versatility significantly, which will be discussed in Sec.\,\ref{subsec:discussion_specific_NBQC}.

In the circuit-specific NBQC, a network consists of $n_{\rm alg}$ qubit and factory components. Unlike circuit-agnostic NBQC, the size of the ring network is determined independently for each qubit component. For the switching network, we use the node-efficient one from the perfect bipartite network or Clos network. In the circuit-specific NBQC, we determine the number of QQ- and QF-links ($M_{i,j}^{QQ}$ and $M_{i}^{QF}$) from the access-frequency explained later. The other network parameters are determined from these values. The number of external ports of the $i$-th qubit component is calculated as $n_{\rm ext}^{(i)} = M_{i}^{QF} + \sum_{j \neq i} M_{i,j}^{QQ}$. We set the number of internal ports to be $n_{\rm int}^{(i)} = \min(n_{\rm ext}^{(i)}, T_{\text{Bell}}/T_{\text{local}})$, and the length of ring network is determined as $n_{\rm ring}^{(i)} = \lceil n_{\rm int}^{(i)}/(d-2) \rceil$. There are $M_i^{QF}$ external ports in the $i$-th factory component, and $T_{\rm magic} / (T_{\text{Bell}} p_{\rm magic})$ magic-state generators are connected to each external port. Therefore, the whole structure can be constructed if $M_{i,j}^{QQ}$ and $M_{i}^{QF}$ are determined.

To determine $M_{i,j}^{QQ}$ and $M_{i}^{QF}$, we characterize the access-frequency profile $\braket{\rm Bias}^{QQ}_{i,j}$ and $\braket{\rm Bias}^{QF}_{i}$ of a target quantum circuit as follows. We first evaluate when each remote operation is executed with the assumption of bottleneck-free behavior. Note that this can be evaluated without a full simulation, as typical quantum algorithms have no program branches except for magic-state gate teleportation. 
We define $c_{i, j}^{(k)}$ and $c_{i}^{(k)}$ as the times when the $k$-th communication event occurs between the $i$-th and the $j$-th qubit component, and between the $i$-th qubit and the $i$-th factory component, respectively. Using them, we can define the access-frequency profiles of the quantum circuit as follows:
\begin{align}
    \braket{\rm Bias}^{QQ}_{i,j} &= \max \left\{ l - k + 1 \mid k < l \text{ and } c_{i, j}^{(l)} - c_{i, j}^{(k)} < T_{\text{Bell}} \right\}, \nonumber \\
    \braket{\rm Bias}^{QF}_{i} &= \max \left\{ l - k + 1 \mid k < l \text{ and } c_{i}^{(l)} - c_{i}^{(k)} < T_{\text{Bell}} \right\}.
\end{align}
In other words, $\braket{\rm Bias}^{QQ}_{i,j}, \braket{\rm Bias}^{QF}_{i}$ represent the maximum number of communication events happening in $T_{\text{Bell}}$ of time duration. From this definition, it is clear that $\braket{\rm Bias}^{QQ}_{i,j} \leq T_{\text{Bell}} / T_{\text{local}}$ and $\braket{\rm Bias}^{QF}_{i} \leq T_{\text{Bell}} / T_{\text{local}}$ are satisfied.

In the circuit-specific network, we choose $M_{i,j}^{QQ} = \braket{\rm Bias}^{QQ}_{i,j}$ and $M_{i}^{QF}=\braket{\rm Bias}^{QF}_{i}$. With this choice, the ring network has a sufficient number of nodes for the frequency of instruction executions. The switching network can support strict-sense non-blocking communication between internal and external ports. The supply rate of magic states is sufficient for each qubit component. The number of external ports is also sufficient for bottleneck-free computation, and the switching network guarantees strict-sense non-blocking connections.

Thus, the overall number of nodes for the circuit-specific bottleneck-free NBQC system becomes $O\left(\sum_{i} \left( \left(\sum_j \braket{\rm Bias}^{QQ}_{i,j}\right)^{\log_s t} + \braket{\rm Bias}^{QF}_{i} \frac{T_{\rm magic}}{T_{\text{Bell}} p_{\rm magic}}\right)\right)$.

\subsubsection{Comparison between Circuit-Agnostic and Circuit-Specific NBQC Designs}
\begin{figure}[t]
    \centering
    \begin{minipage}{0.9\linewidth}
        \centering
        \[ \Qcircuit @C=1em @R=1em {
        \lstick{\ket{Q_0}} & \ctrl{1} & \qw & \targ & \qw & \qw & \ctrl{2} & \targ & \qw & \qw \\
        \lstick{\ket{Q_1}} & \targ & \ctrl{1} & \qw & \qw & \targ & \qw & \qw & \ctrl{2} & \qw \\
        \lstick{\ket{Q_2}} & \ctrl{2} & \targ & \qw & \ctrl{1} & \qw & \targ & \qw & \qw & \qw \\
        \lstick{\ket{Q_3}} & \qw & \qw & \ctrl{-3} & \targ & \qw & \ctrl{1} & \qw & \targ & \qw \\
        \lstick{\ket{Q_4}} & \targ & \qw & \qw & \qw & \ctrl{-3} & \targ & \ctrl{-4} & \qw & \qw
        } \]
        \subcaption{}
        \label{fig:uniform_communication}
    \end{minipage}
    \begin{minipage}{0.9\linewidth}
        \centering
        \[ \Qcircuit @C=1em @R=1em {
        \lstick{\ket{Q_0}} & \ctrl{1} & \ctrl{2} & \ctrl{3} & \ctrl{4} & \ctrl{1} & \ctrl{2} & \ctrl{3} & \ctrl{4} & \qw \\
        \lstick{\ket{Q_1}} & \targ & \qw & \qw & \qw & \targ & \qw & \qw & \qw & \qw \\
        \lstick{\ket{Q_2}} & \qw & \targ & \qw & \qw & \qw & \targ & \qw & \qw & \qw \\
        \lstick{\ket{Q_3}} & \qw & \qw & \targ & \qw & \qw & \qw & \targ & \qw & \qw \\
        \lstick{\ket{Q_4}} & \qw & \qw & \qw & \targ & \qw & \qw & \qw & \targ & \qw
        } \]
        \subcaption{}
        \label{fig:biased_communication}
    \end{minipage}
    \caption{(a) Uniform communication pattern where all pairs of algorithmic qubits have communications with equal frequency. (b)Biased access pattern where all two-qubit operations involve $Q_0$.}
\end{figure}
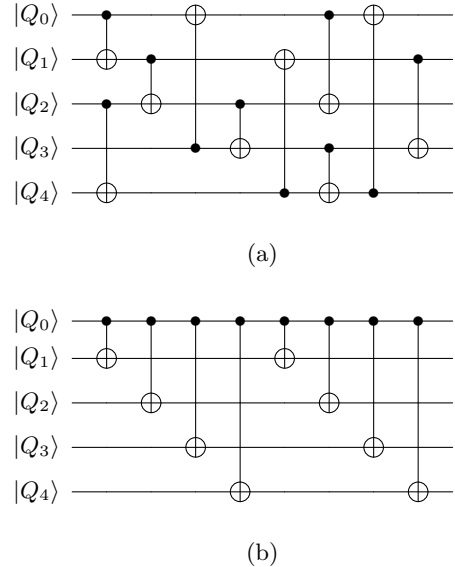

This section compares the scaling of node count required for the circuit-agnostic and circuit-specific NBQC in two access-frequency profiles: the uniform access and the biased access, shown in Figs.~\ref{fig:uniform_communication}, \ref{fig:biased_communication}, respectively.
Note that in this section, we focus on the scaling of qubit components and ignore that of factory components because the dominant term in the number of nodes comes from the switching network inside the qubit component.

In the first example where all algorithmic qubits are uniformly accessed, $O(n_{\rm alg})$ two-qubit operations are performed within $T_{\text{local}}$ of time duration, and the number of two-qubit operations during $T_{\text{Bell}}$ is $O(n_{\rm alg} T_{\text{Bell}}/T_{\text{local}})$. The number of two-qubit gates between the $i$-th and the $j$-th qubit components is $\braket{\rm Bias}^{QQ}_{i,j} = O(T_{\text{Bell}}/(n_{\rm alg} T_{\text{local}}))$.
Thus, the number of nodes for the circuit-specific NBQC is $O\left(n_{\rm alg}\left(T_{\text{Bell}} / T_{\text{local}}\right)^{\log_s t}\right)$, while the circuit-agnostic NBQC requires $O\left( n_{\rm alg}^2 T_{\text{Bell}}/T_{\text{local}} \right)$ nodes.
This result means that the node scaling of the circuit-specific NBQC is $O\left(n_{\rm alg}^{-1} \left(T_{\text{Bell}} / T_{\text{local}}\right)^{(\log_s t) - 1}\right)$ times as many as that of the circuit-agnostic NBQC.

In the next example, where the access pattern is biased to the $0$-th qubit, two-qubit operations between the $0$-th and the $i$-th qubit component are performed sequentially in $T_{\text{local}}$ of time duration, while other pairs of qubit components have no communication. Thus, $\braket{\rm Bias}^{QQ}_{0,i} = T_{\text{Bell}} / (n_{\rm alg} T_{\text{local}})$ holds for any $i$, and $\braket{\rm Bias}^{QQ}_{i,j} = 0$ holds for any $i \neq 0, j \neq 0$.
Using them, the node count of the circuit-specific NBQC is $O\left( \left(T_{\text{Bell}} / T_{\text{local}}\right)^{\log_s t}\right)$, which is smaller than that of the circuit-specific NBQC in the uniform pattern.
In contrast, the number of nodes for the circuit-agnostic NBQC is still $O\left( n_{\rm alg}^2 T_{\text{Bell}}/T_{\text{local}} \right)$.
Thus, the result means that the node scaling of the circuit-specific NBQC is $O\left( n_{\rm alg}^{-2} \left(T_{\text{Bell}} / T_{\text{local}}\right)^{(\log_s t) - 1}\right)$ times as many as that of the circuit-agnostic NBQC.

From these observations, we can conclude that, except in the limited situation where $n_{\rm alg}$ is small and $T_{\text{Bell}} / T_{\text{local}}$ is large, the node scaling of the circuit-specific NBQC is better than that of circuit-agnostic NBQC.
Suppose $(s, t) = (2, 3)$ and $\left(T_{\text{Bell}} / T_{\text{local}}\right) = 1000$ for example. Then, the node scaling of the circuit-specific NBQC becomes more efficient than that of the circuit-agnostic NBQC when $n_{\rm alg} = \left(T_{\text{Bell}} / T_{\text{local}}\right)^{(\log_s t) - 1} \geq 57$ in the uniform-access pattern. In the case of the biased access pattern, it becomes more efficient when $n_{\rm alg} = \sqrt{\left(T_{\text{Bell}} / T_{\text{local}}\right)^{(\log_s t) - 1}} \geq 8$.

\subsection{Working example}
\begin{figure*}[t]
    \centering
    \includegraphics[width=0.9\linewidth]{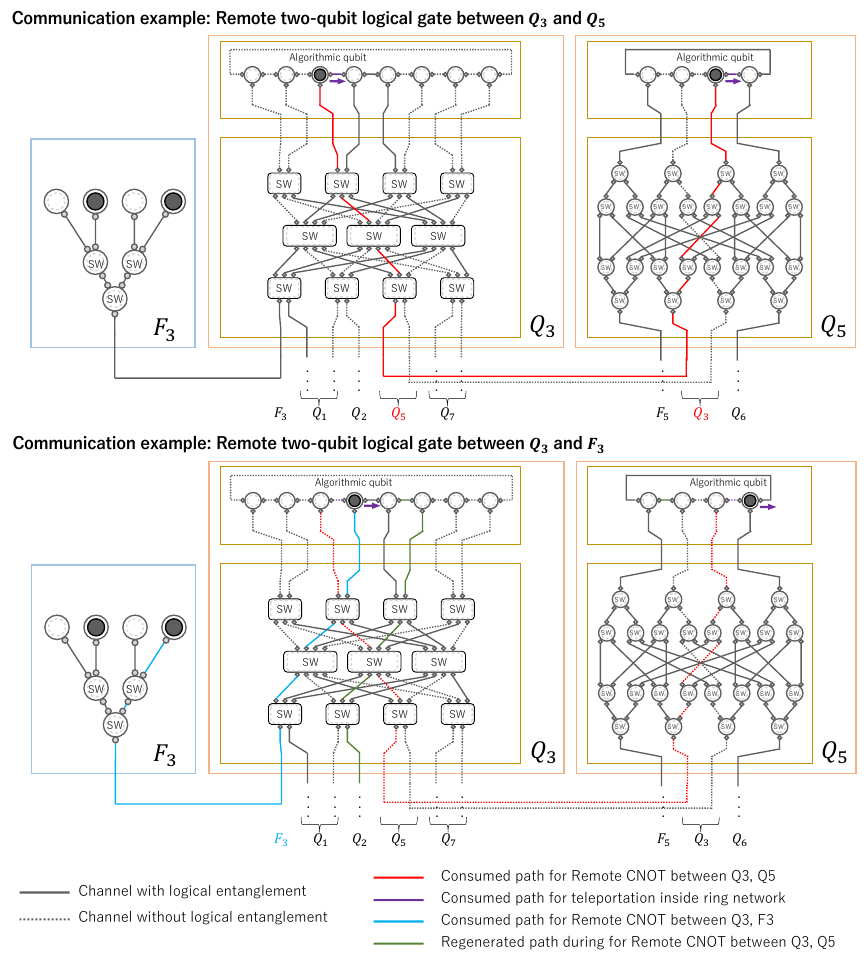}
    \caption{The working example of NBQC. See main text for details.}
    \label{fig:working_example}
\end{figure*}

This section presents a working example of quantum computation based on NBQC with circuit-specific designs. 
Figure\,\ref{fig:working_example} illustrates an example of communication both between qubit components and between qubit and factory components.
Each circle represents a fault-tolerant node that is capable of storing $n_{\rm data}=1$ logical data qubit and communicating with $d=3$ channels, and each switch in the Clos network (the switching network in $Q_3$) is constructed either with a perfect bipartite network or recursively using smaller Clos networks.
Pairs of nodes connected by solid lines possess distilled logical entanglement, whereas those connected by dotted lines do not and are currently regenerating it. 
A filled circle in the qubit component represents a node containing an algorithmic qubit, and filled circles in the factory components correspond to nodes storing magic states.

Suppose we want to perform a logical remote two-qubit gate (e.g., a CNOT or a two-body lattice-surgery operation) between $Q_3$ and $Q_5$. 
We first identify a path between two filled circles in the target qubit components and then consume the logical entanglement along that path. 
In this example, a red path at the top of Fig.\,\ref{fig:working_example} can be used to execute the logical two-qubit gate. 
After the operation, the positions of the algorithmic qubit in the two components are teleported to the nodes on the right side of each ring network. Then, the consumed edges change from solid to dotted lines.

Next, we suppose that we perform a logical remote two-qubit gate between an algorithmic qubit in $Q_3$ and a magic state. We identify a path between the filled circle in the target qubit component and a filled circle in the factory component $F_3$. In this example, a blue path at the bottom of Fig.\,\ref{fig:working_example} can be used for the logical gate. The entangled pairs along the path are consumed in the same manner, and the algorithmic qubit is teleported accordingly.

During these operations, all dotted-line links attempt to regenerate logical entanglement, and some of them will complete the regeneration. Suppose that the green path is a path that has elapsed the longest time since its last consumption. In this case, the entanglement on the green path is expected to have been regenerated. If the length of the ring network and the number of external ports are sufficiently large, this consumption and regeneration process will not stall due to a shortage of distilled entanglement.

\section{NBQC Design with a Limited Number of Nodes}
\label{sec:optimize}
If the number of available nodes is redundant but not sufficient for a bottleneck-free design, it is expected to provide an NBQC network that can satisfy the node-count restriction with minimal overheads of communication according to the access pattern of a target quantum circuit. Here, we show a scheme to heuristically design NBQC networks under the node-count restriction. This starts from the minimum network of circuit-specific NBQCs and iteratively increases the network size to effectively minimize the execution time.

\subsection{Optimization Overview}
Our optimization method involves the following subroutines, each of which is explained in the following subsections:
\begin{enumerate}
    \item \texttt{NBQC Construction} (Sec.~\ref{subroutine:NBQC_construction}): Given the number of QQ-links $M^{QQ}_{i, j}$ and QF-links $M^{QF}_{i}$, construct an NBQC network.
    \item \texttt{Clos Network Optimization} (Sec.~\ref{subroutine:clos_network_optimization}): Given an NBQC network, remove unnecessary nodes from the Clos network without changing the execution time by appropriately assigning an entanglement-swapping path to each remote operation.
    \item \texttt{Bottleneck Identification} (Sec.~\ref{subroutine:bottleneck_identification}): Given an NBQC network, identify the communication bottleneck in the network.
    \item \texttt{Update NBQC Configuration} (Sec.~\ref{subroutine:update_NBQC_configuration}): Given the communication bottleneck, increase the number of QQ-links $M^{QQ}_{i, j}$ and QF-links $M^{QF}_{i}$ to reduce the execution time at the cost of increased node count.
\end{enumerate}
Using these subroutines, our optimization flow can be described as follows:
\begin{itemize}
    \item (Step 1) Initialize the number of QQ- and QF-links as follows. If there is any two-qubit gate between the $i$-th and the $j$-th algorithmic qubit, initialize $M^{QQ}_{i, j} = 1$. Otherwise, set $M^{QQ}_{i, j} = 0$. If the $i$-th algorithmic qubit consumes one or more magic states, initialize $M^{QF}_{i} = 1$. Otherwise, set $M^{QF}_{i} = 0$.
    \item (Step 2) Using the \texttt{NBQC Construction} subroutine, construct an NBQC network from $M_{i,j}^{QQ}$ and $M_i^{QF}$.
    \item (Step 3) Using the \texttt{Clos Network Optimization} subroutine, remove unnecessary nodes from the network.
    \item (Step 4) Using the \texttt{Bottleneck Identification} subroutine, find the communication bottleneck of the current NBQC design.
    \item (Step 5) If there is no bottleneck found or there remain no redundant nodes, return the current NBQC network. Otherwise, use the \texttt{Update NBQC Configuration} subroutine to update $M^{QQ}_{i, j}$ and $M^{QF}_{i}$ and go back to Step 2.
\end{itemize}
With the above procedure, we can obtain a circuit-specific NBQC network optimized for execution time while satisfying the upper limit of the total number of nodes.
Note that our construction does not guarantee the optimal solution and may require many iterations when the number of available nodes is large. To improve the optimization procedure, it is possible to use other heuristic algorithms such as beam search and simulated annealing. The exploration of faster network optimization is left as future work.

\subsection{Subroutine: NBQC Construction}
\label{subroutine:NBQC_construction}
This subroutine receives the number of QQ-links and QF-links, represented as $M^{QQ}_{i, j}$ and $M^{QF}_{i}$, and returns an NBQC network based on these values.
For the $i$-th qubit component, the number of external ports is chosen as $n_{\rm ext}^{(i)} = M^{QF}_{i} + \sum_{j \neq i} M^{QQ}_{i, j}$ and the number of internal ports as $n_{\rm int}^{(i)} = \min\left( T_{\text{Bell}}/T_{\text{local}}, n_{\rm ext} \right)$. They are connected with a node-efficient one among the perfect bipartite network and the Clos network. If we use the Clos network, the parameter $(s,t)$ is set to $t=2s-1$, and $s$ is optimized to minimize the total number of nodes.

For the $i$-th factory component, we design it to have $M^{QF}_{i}$ external ports. Note that $M^{QF}_{i} = 0$ means that the $i$-th factory component is empty. For each external port, we prepare $T_{\rm magic} / (T_{\text{Bell}} p_{\rm magic})$ magic-state generators and connect with the external port via a tree network. If $T_{\rm magic} / (T_{\text{Bell}} p_{\rm magic})$ magic-state generators are too many, the number of magic state generators is reduced so that it does not affect the execution time.

\subsection{Subroutine: Clos Network Optimization}
\label{subroutine:clos_network_optimization}
The aim of this subroutine is to reduce the number of nodes in the Clos networks without the penalty of execution times. The Clos networks in the output of subroutine \texttt{NBQC Construction} consist of $N_{\rm Clos}(n_{\rm int}, n_{\rm ext}; d)$ nodes. While this guarantees strict-sense non-blocking communication for any profile, there is room to reduce the number of nodes in Clos networks without increasing execution time if we can assume a specific target quantum circuit. This can be rephrased as follows: let a graph of the switching network in the target qubit component be $G$, and the execution time of a target quantum circuit with a graph be $T(G)$. Then, our aim is to \textit{minimize the number of nodes in $G$ so that $T(G) = T(G_0)$, where $G_0$ is a strict-sense non-blocking switching network}.

This optimization problem is computationally too expensive since investigating all possible graphs $G$ is difficult. Thus, we approximately solve it by limiting the target of node reduction to the number of middle switches. If each middle-stage switch consists of Clos networks, we repeat this procedure iteratively. Let $T_{\rm reduce}(R)$ be an execution time under reduced Clos networks with $R$ middle-stage switches. Then, the optimization problem is converted to \textit{minimize the number of middle switches $R$ under constraint of $T_{\rm reduce}(R) = T(G_0)$}. 

The condition $T_{\rm reduce}(R) = T(G_0)$ is rephrased as follows. Let $\tilde{t}^{(k)}_{i}$ be the timing when the $k$-th remote logical operation acting on the $i$-th qubit component is executed using a strict-sense non-blocking switching network. Let $\tilde{a}^{(k)}_{i} \in [1...\ceil{n_{\rm int} / s}]$, $\tilde{b}^{(k)}_{i} \in [1...R]$, and $\tilde{c}^{(k)}_{i} \in [1...\ceil{n_{\rm ext} / s}]$ are the switch indices of the left, middle, and right stage used in that operation, respectively. Here, $\tilde{a}^{(k)}_{i}$ and $\tilde{c}^{(k)}_{i}$ are determined by the wiring between internal (external) ports to the switches in the left (right) stage, and $\tilde{b}^{(k)}_{i} \in [1...R]$ can be chosen for each instruction.
The execution time of a reduced Clos network becomes the same as bottleneck-free cases when there is a choice of $\tilde{a}^{(k)}_{i}, \tilde{b}^{(k)}_{i}, \tilde{c}^{(k)}_{i}$ such that no pair of instructions, $k$ and $l$ satisfies the following two conditions at the same time.
\begin{enumerate}
    \item Two instructions are executed within $T_{\text{Bell}}$: $|\tilde{t}^{(l)}_{i} - \tilde{t}^{(k)}_{i}| < T_{\text{Bell}}$.
    \item Two instructions share the same middle switch and another switch: $\tilde{b}^{(k)}_{i} = \tilde{b}^{(l)}_{i}$ and ($\tilde{a}^{(k)}_{i} = \tilde{a}^{(l)}_{i}$ or $\tilde{c}^{(k)}_{i} = \tilde{c}^{(l)}_{i}$).
\end{enumerate}

\begin{figure*}[t]
    \centering
    \begin{minipage}{0.9\linewidth}
        \centering
        \includegraphics[width=\linewidth]{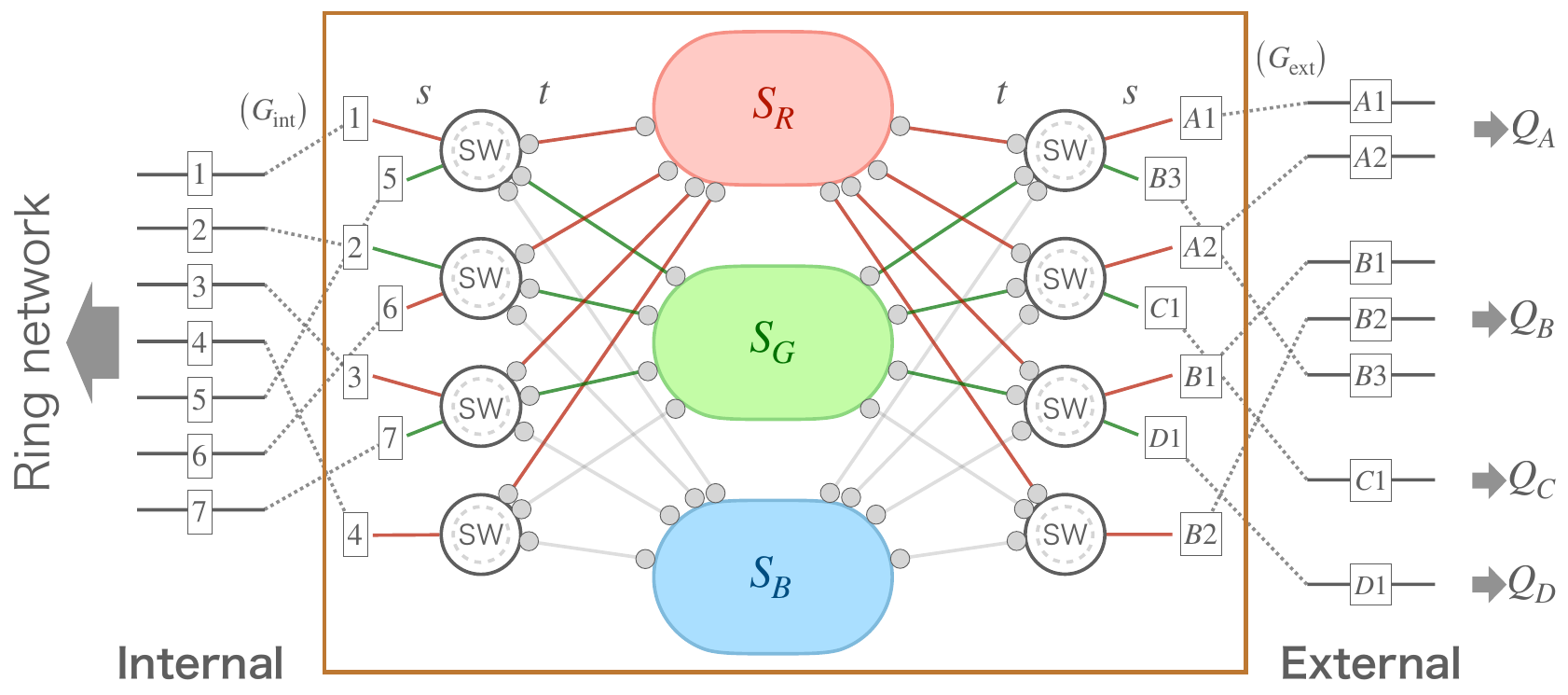}
        \subcaption{}
        \label{fig:clos_network_optimization_grouping}
    \end{minipage}
    \begin{minipage}{0.7\linewidth}
        \centering
        \includegraphics[width=\linewidth]{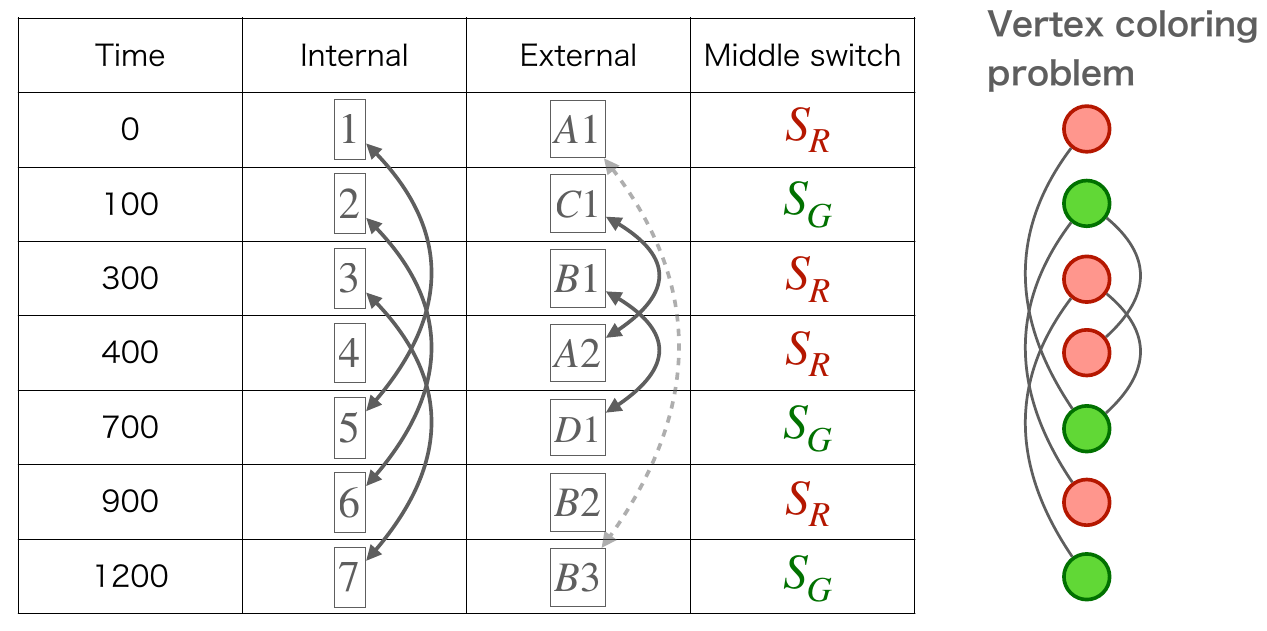}
        \subcaption{}
        \label{fig:clos_network_optimization_coloring}
    \end{minipage}
    \caption{Overview of Clos network optimization. In this example, we set $n_{\rm int} = n_{\rm ext} = 7, (s, t) = (2, 3)$, and $T_{\text{Bell}} = 1000$. The external ports are connected to other qubit components $Q_A, Q_B, Q_C$, and $Q_D$. (a) At the first step of optimization, we design mappings from ports to switches. (b) For each communication event, we determine which middle switch to use while maximizing the number of unused switches. This problem can be formulated as a vertex coloring problem, which uses three colors: $S_R, S_G$, and $S_B$ in this example. Note that the middle switch $S_B$ can be removed since no communication event uses it.}
    \label{fig:clos_network_optimization}
\end{figure*}

To simplify the problem, we fix $\tilde{a}^{(l)}_{i}$ and $\tilde{c}^{(l)}_{i}$ by connecting internal (external) ports to switches in the left (right) stage cyclically, i.e., the $i$-th internal port is connected to the $(i \mod \ceil{n_{\rm int} / s})$-th switch in the left stage, and the $i$-th external port to the $(i \mod \ceil{n_{\rm ext} / s})$-th switch in the right stage, as shown in Fig.~\ref{fig:clos_network_optimization_grouping}. This cyclic assignment is chosen because the internal ports and each group of external ports connected to the same component are used in a cyclic manner, and we can heuristically expect that cyclic assignment would typically satisfy $\tilde{a}^{(k)}_{i} \neq \tilde{a}^{(l)}_{i}$ or $\tilde{c}^{(k)}_{i} \neq \tilde{c}^{(l)}_{i}$ for two subsequent instructions.

Then, residual variables are $\tilde{b}^{(k)}_{i} \in [1..R]$, and their optimal assignment to minimize $R$ can be converted to a vertex coloring problem as follows. We consider a graph where each vertex represents an instruction acting on the $i$-th qubit component. If two instructions satisfy $|\tilde{t}^{(l)}_{i} - \tilde{t}^{(k)}_{i}| < T_{\text{Bell}}$ and ($\tilde{a}^{(k)}_{i} = \tilde{a}^{(l)}_{i}$ or $\tilde{c}^{(k)}_{i} = \tilde{c}^{(l)}_{i}$), they are connected by an edge. The color corresponds to the choice of middle-stage switch $\tilde{b}^{(k)}_{i} \in [1..R]$~(see Fig.~\ref{fig:clos_network_optimization_coloring}). If there is a node coloring with $R$ colors such that no connected nodes have the same color, all the instructions are scheduled without communication-time overheads and $T_{\rm reduce}(R) = T(G_0)$ is satisfied. Otherwise, instructions corresponding to a pair of connected nodes would satisfy all the conditions listed above.

Although the graph coloring problem is NP-hard in general, we can use greedy algorithms, such as the Welsh-Powell algorithm~\cite{welsh_powell}, to obtain a near-optimal solution. After that, we can eliminate switches and unused channels in the middle stage without affecting the total execution time. 

\subsection{Subroutine: Bottleneck Identification}
\label{subroutine:bottleneck_identification}
This subroutine receives the NBQC network and evaluates the access timing of instructions in a target quantum circuit. From this evaluation, we define a value named communication bottleneck for each operation as follows.
For the $k$-th remote operation, it might need to wait for the Bell generation due to the lack of a QQ-link or a QF-link. If the $k$-th remote operation connects the $i_k$-th and the $j_k$-th qubit component, $w^{QQ (k)}_{i_k, j_k}$ is defined as the duration of time from when the operation is requested until it becomes ready. If the $k$-th remote operation consumes a magic state and acts on the $i_k$-th qubit component, $w^{QF (k)}_{i_k}$ is defined similarly. For other QQ-links and QF-links, $w^{QQ (k)}_{i, j}$ and $w^{QF (k)}_{i}$ are set to zero.

\subsection{Subroutine: Update NBQC Configuration}
\label{subroutine:update_NBQC_configuration}
Given communication bottlenecks $w^{QQ (k)}_{i, j}$, $w^{QF (k)}_{i}$ defined in the previous section, this subroutine returns which pair of components should be selected to increase the number of QQ- or QF-links between them. Our task is to estimate the effect of increasing $M^{QQ}_{i, j}$ or $M^{QF}_{i}$ on the total number of nodes and the execution time.

First, let us estimate the increase in the number of nodes. If we change $M^{QQ}_{i, j}$ to $M^{QQ}_{i, j} + 1$, it requires an additional external port of the $i$-th and the $j$-th qubit component, extends their ring networks, and increases the size of the switching network. Similarly, changing $M^{QF}_{i}$ to $M^{QF}_{i} + 1$ requires additional external ports of the $i$-th qubit component and the $i$-th factory component, as well as the number of magic-state generators along with it. The increase in node count can be estimated by reconstructing an NBQC design with a new configuration and calculating the difference. We denote the increase in node count as $\Delta N^{\text{total}, QQ}_{i, j}$, $\Delta N^{\text{total}, QF}_{i}$ for QQ-links and QF-links, respectively.

Next, we estimate the total improvement in the execution time by increasing the number of QQ-links or QF-links. Assume that the number of QQ-links is changed from $M^{QQ}_{i, j}$ to $M^{QQ}_{i, j} + 1$. Then, the average time for generating a Bell pair is improved from $T_{\text{Bell}} / M^{QQ}_{i, j}$ to $T_{\text{Bell}} / (M^{QQ}_{i, j} + 1)$, which effectively improves the latency of generating a Bell pair by $T_{\text{Bell}} \left( \left(M^{QQ}_{i, j}\right)^{-1} - \left(M^{QQ}_{i, j} + 1\right)^{-1} \right)$. Thus, given $w^{QQ (k)}_{i, j}$ for $k$-th operation consuming a Bell pair between the $i$-th and the $j$-th qubit component, the improvement in the execution time can be approximately estimated as $\Delta w^{QQ (k)}_{i, j}:= \min\left( w^{QQ (k)}_{i, j}, T_{\text{Bell}} \left( \left(M^{QQ}_{i, j}\right)^{-1} - \left(M^{QQ}_{i, j} + 1\right)^{-1} \right) \right)$. Same discussion can be applied to QF-links and we can obtain $\Delta w^{QF (k)}_{i}:= \min\left( w^{QF (k)}_{i}, T_{\text{Bell}} \left( \left(M^{QF}_{i}\right)^{-1} - \left(M^{QF}_{i} + 1\right)^{-1} \right) \right)$ as well.

Using these metrics, one can choose the pair of components which maximizes $(\sum_k \Delta w^{QQ (k)}_{i, j}) / \Delta N^{\text{total}, QQ}_{i, j}$ or $(\sum_k \Delta w^{QF (k)}_{i}) / \Delta N^{\text{total}, QF}_{i}$ as the one which has the largest improvement per extra node.

\section{Numerical evaluation}
\label{sec:evaluation}
\subsection{Evaluation setting}
In this section, we numerically evaluate the execution time and the number of required nodes. During our simulation, we assume that each node can store a single algorithmic qubit ($n_{\rm node}=1$) and the number of channels per node ($= d$) varies in terms of the evaluation. In constructing Clos networks, we used $(s, t) = (2, 3)$ as the design parameters. In our numerical analysis, while $n_{\rm int}$ can be smaller than $n_{\rm ext}$, we set $n_{\rm int} = n_{\rm ext}$ to simplify the construction of Clos networks.

We assumed logical qubits are encoded in surface codes, and we refer to the latency of each logical operation in the case of surface codes as summarized in Table \ref{tab:operation_latency}.
The time unit refers to the duration required to repeat syndrome measurements as many times as the code distance, which is equal to a clock in Ref.\,\cite{litinski2019game} and code beat in Ref.\,\cite{yoshioka2024hunting}. Note that while we refer to the parameters of surface codes used in Ref.\cite{beverland2022assessing}, the framework of NBQC can be applied to general cases, as discussed in Sec.\,\ref{sec:discussion}.
For magic state generation, we assume that the magic state is prepared via a magic-state cultivation protocol~\cite{gidney2024magic} and its fidelity is sufficient for logical operations.
To the best of our knowledge, there is no consensus on the experimentally demonstrated generation time of logical Bell pairs. We heuristically assume that each Bell generation takes 1,000 time units.

\begin{table*}[t]
    \centering
    \begin{tabular}{l|c|c}
        \hline
        Operation   & Latency & Note \\
        \hline
        \hline
        Initialization in $X,Z$ basis & 0 time unit &  \\
        \hline
        Destructive measurement  in $X,Z$ basis & 0 time unit &  \\
        \hline
        $H$ gate & 3 time unit &  \\
        \hline
        $S, S^{\dagger}$ gate & 2 time unit & \\
        \hline
        Lattice surgery & 1 time unit & \\
        \hline
        CNOT gate & 2 time unit & \\
        \hline
        Magic state generation & 2 time unit & succeed with probability 0.01 \\
        \hline
        logical entanglement generation & 1000 time unit &  \\
        \hline
        quantum state teleportation & 1 time unit & consume logical entanglement \\
        \hline
    \end{tabular}
    \caption{Latency for each logical operation.}
    \label{tab:operation_latency}
\end{table*}

When we translate quantum circuits to a sequence of instructions, we use {\tt pygridsynth}~\cite{pygridsynth,ross2014optimal} to convert them into Clifford+T form, and the application of $T$ gates is decomposed as the combination of lattice surgery and the feedforward of Clifford gates. While the Clifford gates after magic-state teleportation are applied with probability $0.5$, we assume they are always skipped for simplicity.

\subsection{Benchmark Circuits}
We evaluate the performance of NBQC using various quantum circuits in QASMBench~\cite{li2022qasmbench}. From QASMBench, we choose \texttt{adder\_n28}, \texttt{multiplier\_n15}, \texttt{qram\_n20} to demonstrate that NBQC works properly regardless of the type of application.
In addition, we evaluate NBQC using a \texttt{SELECT} circuit, which are bottleneck subroutines in the current state-of-the-art quantum phase estimation~\cite{yoshioka2024hunting,low2019hamiltonian,babbush2018encoding,harrigan2024expressing}. See Refs.\,\cite{babbush2018encoding,yoshioka2024hunting} for the detailed definitions of the \texttt{SELECT} circuits.
As the construction of \texttt{SELECT} depends on the target Hamiltonian of quantum phase estimation, we use the Heisenberg model with the 2D cylinder topology with size $6 \times 6$, $8 \times 8$, and $10 \times 10$. According to Ref.\,\cite{yoshioka2024hunting}, quantum advantage is expected from the size of $10 \times 10$ or larger.

\subsection{Trade-offs between Execution Time and Node Count}
\begin{figure*}[t]
    \centering
    \includegraphics[width=0.9\linewidth]{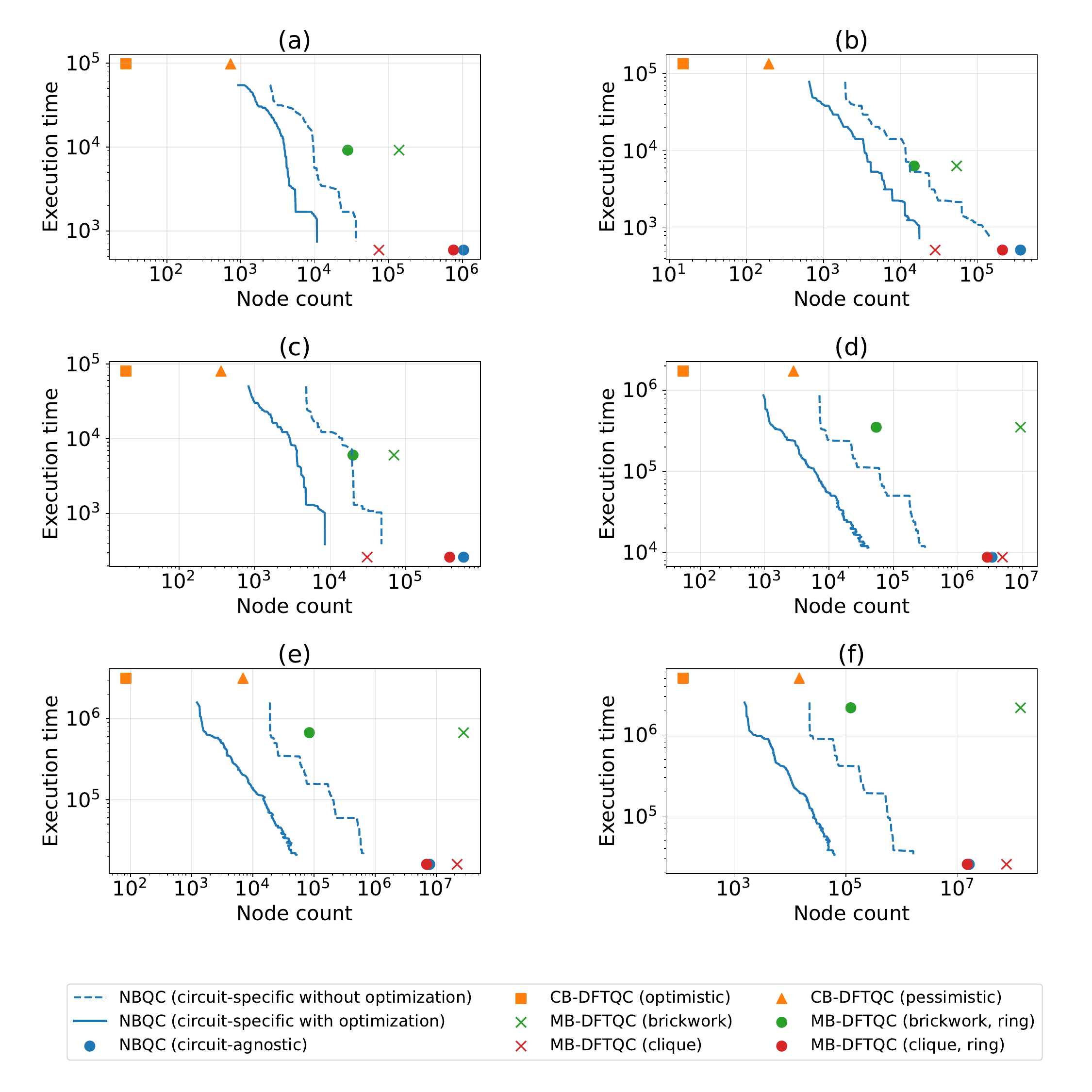}
    \caption{
        The trade-off relation between the execution time and node count. We choose (a) \texttt{adder\_n28}, (b) \texttt{multiplier\_n15}, and (c) \texttt{qram\_n20} from QASMBench, and the \texttt{SELECT} circuits for quantum phase estimation algorithm with the Heisenberg model on the 2D cylinder topologies of (d) $6 \times 6$, (e) $8 \times 8$, (f) $10 \times 10$ particles, respectively.
        The solid lines represent those of NBQC, and the dotted lines represent the result of NBQC without the \texttt{Clos Network Optimization} subroutine presented in Sec.\,\ref{sec:optimize}. We also plot the execution time and node count for CB-DFTQC and MB-DFTQC for comparison. The number of channels per node ($=d$) is fixed to 3 in this evaluation.
    }
    \label{fig:result_QASMBench_SELECT}
\end{figure*}

Figure~\ref{fig:result_QASMBench_SELECT} shows the trade-off relations between the execution time and node count in benchmark circuits. 
For the evaluation of NBQC, we assumed the number of channels is $d=3$. The performance of a bottleneck-free circuit-agnostic NBQC is shown as a blue circle. The performance of circuit-specific NBQCs built for a given node count is shown as blue lines. The solid and broken lines correspond to the performance with and without the call of the subroutine \texttt{Clos Network Optimization} in Sec.\,\ref{sec:optimize}. In the evaluation of circuit-specific NBQC, we build NBQC networks for each target quantum circuit.

As a reference, we also evaluate the execution time and node count of CB-DFTQC and MB-DFTQCs. As their exact evaluation and detailed optimization are out of the scope of this paper, we evaluated their performance in a simplified manner as follows. In the evaluation of CB-DFTQC and MB-DFTQC, we ignore execution time and node count for generating magic states. For CB-DFTQC, we plotted two points with different node-counting methods, optimistic (yellow square) and pessimistic (yellow triangle). For an optimistic point, we assume every node has $d=n_{\rm alg}$ channels, i.e., an all-to-all connection is available. Thus, the number of nodes is equal to $n_{\rm alg}$. For a pessimistic point, we assume $d=3$ and assume each algorithmic qubit needs a tree network with its leaves $n_{\rm alg}-1$ to support an all-to-all network, which requires $n_{\rm alg} \times N_{\rm tree}(n_{\rm alg}-1; d) = n_{\rm alg}(n_{\rm alg}-2)$ nodes. Thus, the actual node count is expected to be an intermediate between these two points. In the evaluation of MB-DFTQC, we checked the performance for two types of cluster states, Brickwork and Clique cluster states, which are shown as green and red points, respectively. We also evaluate cases where the whole cluster state is generated at the beginning (cross marker), or the cluster state is generated and consumed at runtime on a ring-shaped network (circle).

Note that CB-DFTQC with pessimistic setting (yellow triangle) and MB-DFTQC with clique ring (red circle) are almost equal to circuit-agnostic NBQC settings with $n_{\rm ring}=1$ (ring not used) and $n_{\rm ring}=T_{\text{Bell}}/T_{\text{local}}$ (bottleneck-free) without \texttt{Clos Network Optimization}, respectively. A major difference is that CB-DFTQC and MB-DFTQC ignore the time and nodes for magic-state generations, while NBQC treats them faithfully with factory components.

We can see that circuit-agnostic NBQC can achieve algorithmic execution time. Also, circuit-specific NBQCs successfully trade the number of nodes and execution time. 
In most examples, we observe that the scaling of execution time is almost inversely proportional to the node count. Instances of \texttt{adder\_n28} and \texttt{qram\_n20} show relatively better scaling to other cases, approximately 100 times speed up using 10 times more nodes. We guess this is because the bottleneck is concentrated on a specific part of the network, and the execution time can be improved with a small addition of nodes.
Note that we observed some jumps in node count at several points. This jump occurs because the number of nodes in the Clos network significantly increases when the number of internal or external ports increases from $s^k$ to $s^k+1$.

It should be noted that the number of nodes, as well as such an increase, can be suppressed without compromising the execution time by the \texttt{Clos Network Optimization} subroutine discussed in Sec.\,\ref{sec:optimize}. Comparing the solid lines to the dotted lines, we can see that NBQC systems without Clos network optimization require 2 to 10 times as many nodes as optimized NBQC systems, which indicates that most nodes in Clos networks are not being used.

When we compare NBQC with CB-FTQC and MB-FTQC, the performance of NBQC with \texttt{Clos Network Optimization} achieves shorter execution times than CB-FTQC, and a smaller number of nodes than MB-FTQC. Note that a circuit-agnostic NBQC requires a larger node count than the MB-DFTQC because the MB-DFTQC ignores the node count for magic-state generations. We believe they show almost the same performance if they are evaluated with the same settings.

\subsection{Effects of the Number of Channels}
\begin{figure*}[t]
    \centering
    \begin{minipage}{0.45\linewidth}
        \centering
        \includegraphics[width=\linewidth]{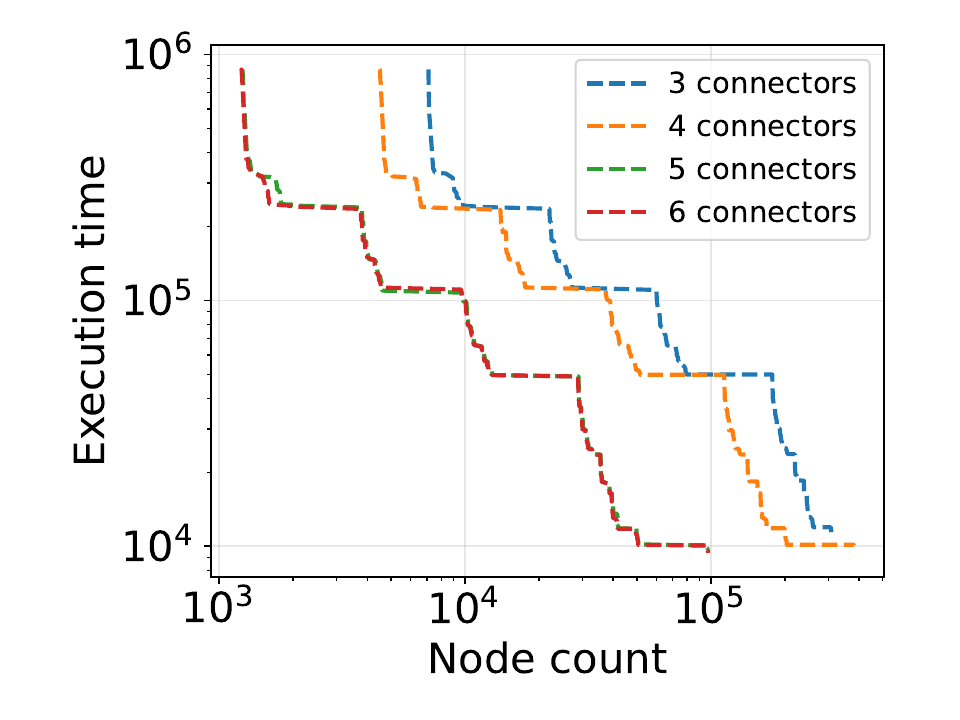}
        \subcaption{}
        \label{fig:result_num_channels_naive}
    \end{minipage}
    \begin{minipage}{0.45\linewidth}
        \centering
        \includegraphics[width=\linewidth]{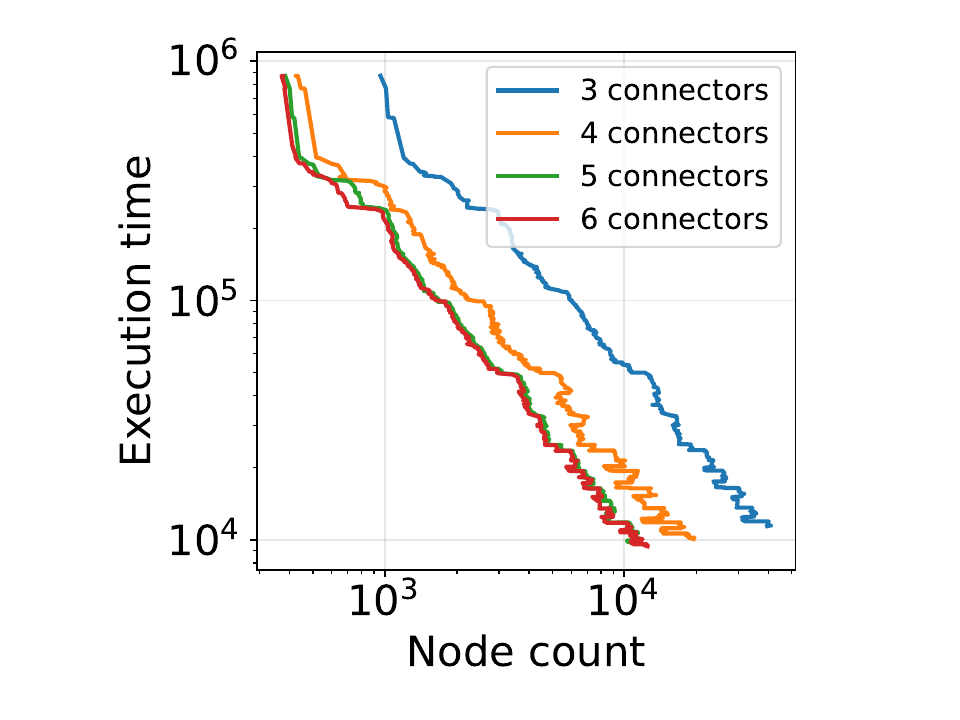}
        \subcaption{}
        \label{fig:result_num_channels_reduced}
    \end{minipage}
    \caption{The effect of the number of channels per node. We use the \texttt{SELECT} circuits for the quantum phase estimation algorithm with the system size of $6 \times 6$ from our benchmark, and estimate the execution time and node count. (a) shows the results without the \textit{Clos Network Optimization} subroutine, and (b) displays the results after optimization.}
    \label{fig:result_num_channels}
\end{figure*}

Next, we investigate the effect of the number of channels per node $d$ on the total node count. In this evaluation, we vary the number of channels for $d=3, 4, 5, 6$ and plot the node count and execution time. The other settings are the same as the last section. Figure~\ref{fig:result_num_channels} shows the results, where the Figs.\,\ref{fig:result_num_channels_naive} and \ref{fig:result_num_channels_reduced} correspond to the performance without and with \texttt{Clos Network Optimization} subroutine, respectively.

Our results indicate that the difference in the number of channels per node has a large impact on the node count.
Without \texttt{Clos Network Optimization}, the most time-efficient design, the $d=3$ design requires $5.94$ times as many nodes as the $d=6$ design. On the other hand, when we enable the optimization, it takes $3.29$ times as many nodes.
Such a difference comes from the overhead of constructing switches in Clos networks.
For example, we can construct a switch with $(s, t) = (2, 3)$ using only one node with $d \geq 5$, while it takes $N_{\rm bipartite}(2, 3; 3) = 7$ nodes when $d=3$.
Despite the large difference between $d=3$ and $d \geq 5$, it is also shown that the increase from $d=5$ to $d=6$ has little improvement in the number of nodes. This is because $d=5$ is sufficient to construct switches with $(s, t)=(2, 3)$, which corresponds to $N_{\rm bipartite}(2, 3; 5) = N_{\rm bipartite}(2, 3; 6) = 1$.

Interestingly, we observe that the behavior of $d=4$ is similar to that of $d=3$ before applying the \texttt{Clos Network Optimization} subroutine, while it is similar to that of $d=5$ after optimization. We guess this is due to the reduction of unnecessary communication paths in Clos networks. Each switch in Clos networks requires $N_{\rm bipartite}(2, 3; 4) = 5$ nodes when $d=4$. However, if at least one edge is removed during Clos network optimization, the switch can be implemented using only one $d=4$ node.

\section{Discussion}
\label{sec:discussion}
This section discusses the qualitative differences between NBQC and CB-DFTQC/MB-DFTQC, and provides an intuitive explanation of why NBQC can achieve a short execution time with a few nodes. Also, we provide several potential extensions of the NBQC framework.

\subsection{Difference between NBQC and CB-DFTQC}
\begin{figure*}[t]
    \centering
    \begin{minipage}{1.0\linewidth}
        \centering
        \includegraphics[width=\linewidth]{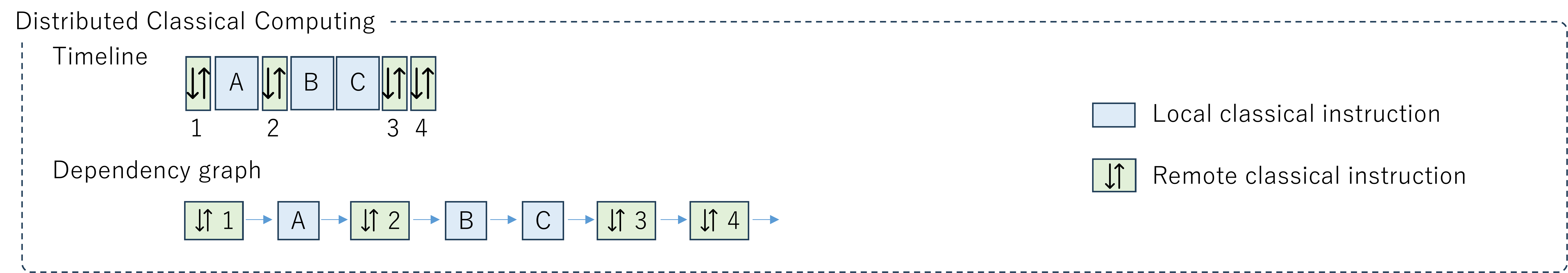}
        \subcaption{Timeline and dependency graph of classical distributed computing.} 
        \label{fig:timeline_classical}
    \end{minipage}
    \begin{minipage}{1.0\linewidth}
        \centering
        \includegraphics[width=\linewidth]{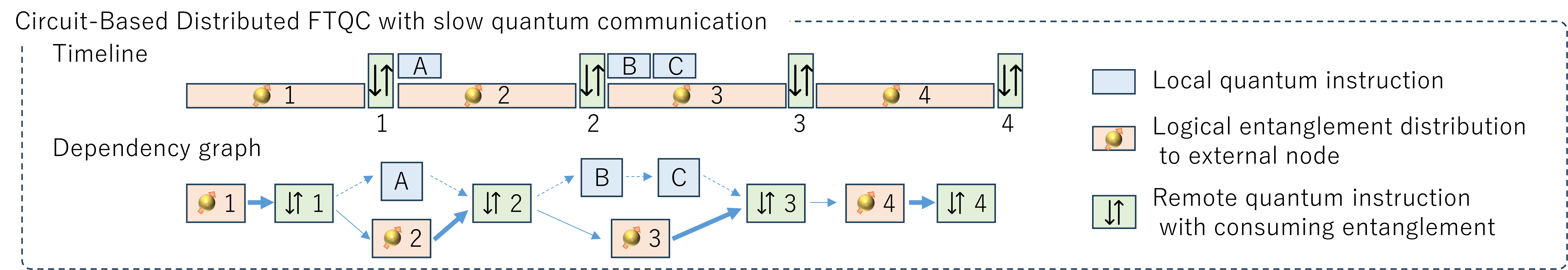}
        \subcaption{Timeline and dependency graph of CBQC.} 
        \label{fig:timeline_cbqc}
    \end{minipage}
    \begin{minipage}{1.0\linewidth}
        \centering
        \includegraphics[width=\linewidth]{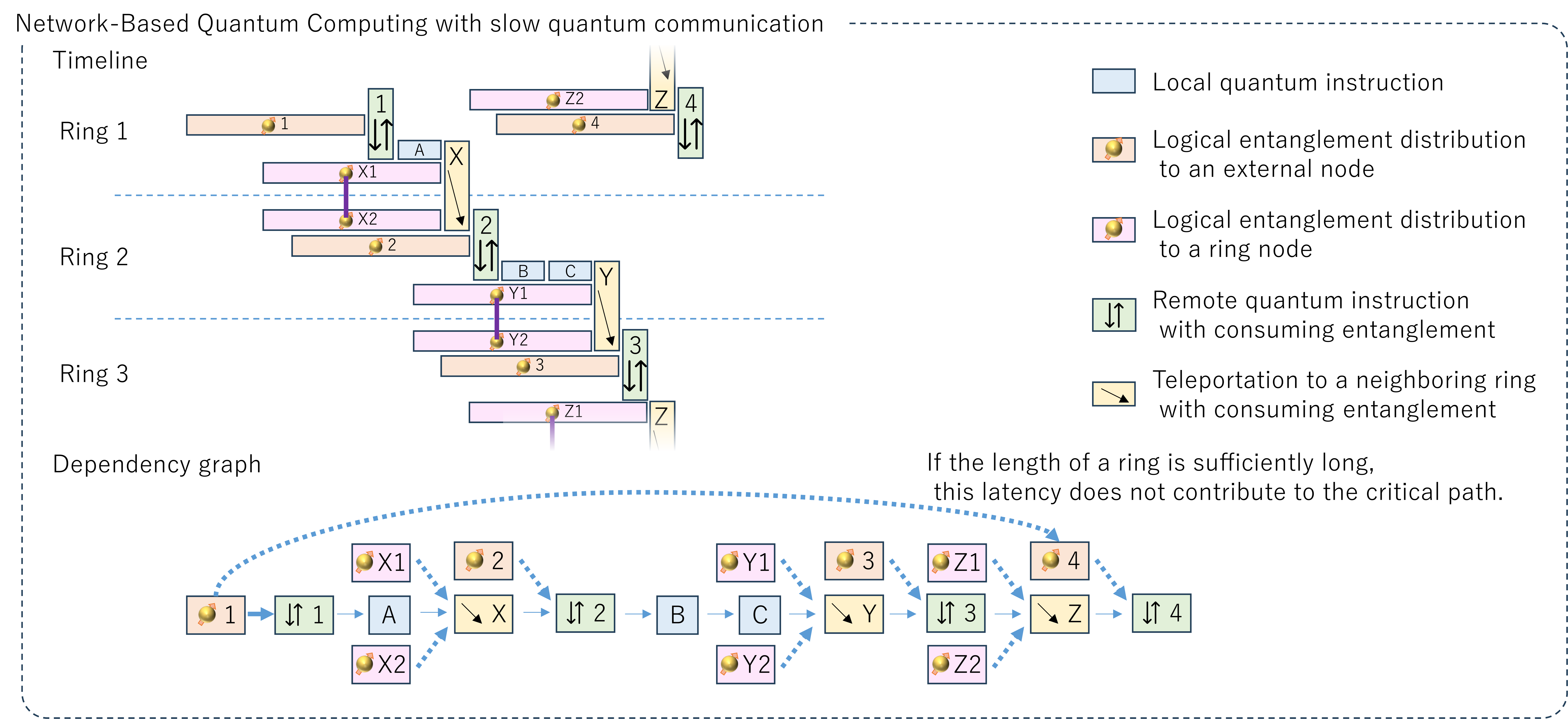}
        \subcaption{Timeline and dependency graph of NBQC.}
        \label{fig:timeline_nbqc}
    \end{minipage}
    \caption{The timeline and dependency graphs for classical distributed computing, circuit-based quantum computing~(CBQC), and network-based quantum computing~(NBQC). The solid line in the dependency graph corresponds to the critical path.}
\end{figure*}

This section explains how NBQC leverages the property of quantum communication to achieve algorithmic execution time from the aspects of program analysis.
Suppose we need to perform a sequence of classical instructions, where each depends on the output of the previous instruction. Since these instructions are sequential, there is no way to speed up the execution time.
Figure~\ref{fig:timeline_classical} shows the timeline and dependency graph of an example process. The timeline is a list of instructions where the horizontal axis represents time, each box represents an instruction, and the width of each box indicates its latency. A dependency graph is a directed acyclic graph in which each node corresponds to an instruction, and there is an edge from A to B if B must be executed after A finishes. The execution time can be characterized by the longest path in the dependency graph, known as the critical path. The edges on the critical path are drawn with solid lines, and the others with dotted lines.

We can consider a quantum variant of this problem, i.e., replacing the classical instructions with quantum ones while keeping the same dependencies. Here, we assume that quantum communication is much slower than classical communication and local quantum instructions. There is a critical difference between classical and quantum communication. Quantum communication can be divided into two steps: the distribution of logical entanglement and the remote operations that consume it. Since the latter requires only local operations and classical communication, it is much faster than the former. The first step is time-consuming, but we can start it before completing the previous instructions.
Figure~\ref{fig:timeline_cbqc} clearly illustrates that the length of the critical path can be reduced by parallelizing the logical entanglement generation and local operations. On the other hand, if quantum communication is very slow, the execution time is still dominated by the time required to generate logical entanglement. To illustrate the number of communication latencies on the critical paths, we draw slow instructions with wide arrows in the dependency graph, and there are four slow communications on the critical path in this example.

By leveraging the property of quantum communication, we can further reduce the critical path by adding a redundant node, which is a core part of NBQC. The timeline and dependency graph for NBQC are shown in Fig.~\ref{fig:timeline_nbqc}. If we are allowed to use ring networks to store algorithmic qubits, we can teleport them to the neighboring node without waiting for the subsequent entanglement to become ready. If the rings are sufficiently long, the latency of logical entanglement generation can be concealed from the critical path except for the first one. While there is a penalty due to the latency of teleporting algorithmic qubits to a neighboring node, it contributes to the execution time only by a constant factor. Thus, we can conclude that this technique for concealing communication latency is quantum-specific.

The necessary length of each ring depends on the ratio between local operations and remote operations acting on the stored algorithmic qubits. If the algorithmic qubits rarely request remote quantum instructions, a short ring length is sufficient to minimize the critical path. This means that if there is a bias in access frequencies in the quantum program, the length of its critical path can be reduced with a modest overhead in node count. As explained in the next section, NBQC is designed to exploit this property. While MBQC can partially utilize the benefit of this property, it is not designed to fully exploit it.

One might wonder whether actual quantum programs exhibit such biased access patterns in realistic applications. Ref.~\cite{kobori2025lsqca} recently revealed that the bottleneck subroutines of state-of-the-art quantum algorithms with exponential speed-ups, such as the SELECT operations in Quantum Phase Estimation~\cite{babbush2018encoding,yoshioka2024hunting}, show strong biases in their compiled programs. Even if every algorithmic qubit is accessed uniformly, we can insert SWAP operations to concentrate the operands of quantum instructions onto a small subset of algorithmic qubits if there is a temporal correlation in the access pattern, i.e., access locality.

\begin{figure*}[t]
    \centering
    \includegraphics[width=0.9\linewidth]{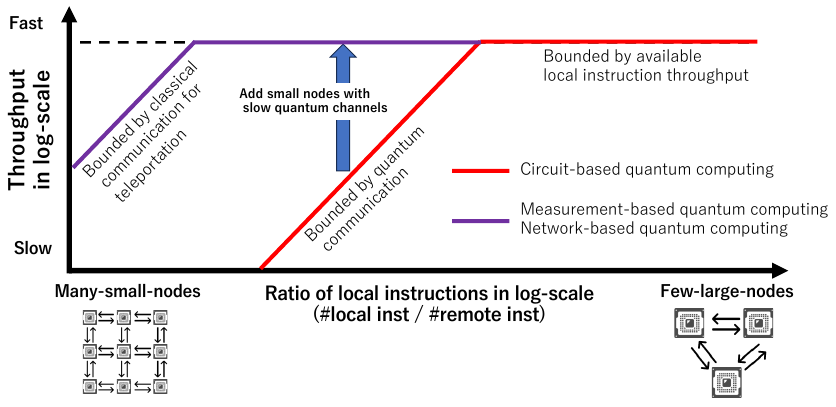}
    \caption{A Roofline model considering the idea of MBQC and NBQC.}
    \label{fig:roofline}
\end{figure*}

While the above fact may be implicitly known in the field of MBQC, its implication may be non-trivial in the context of parallel computing. In parallel computing, the execution time of a computer can be estimated using the ratio of communication to computation. If we divide a single large node into several smaller nodes, the number of local operations per node decreases while communication overhead increases. Once communication becomes the bottleneck, the overall performance is limited by the communication speed, and such a situation is known as communication-bounded. This performance projection methodology is known as the Roofline model~\cite{williams2009roofline} (Fig.\,\ref{fig:roofline}) and can be applied to various types of computational models. The above fact about quantum communication indicates that the Roofline model does not hold for quantum communication and quantum computation in DFTQC, since the major latency of quantum communication can always be removed from the critical path through parallelization.
While the Roofline model holds for quantum computation combined with \textit{classical communication}, communication-bounded situations are unlikely to occur in typical applications, as classical communication is much faster than the latency of fault-tolerant quantum instructions.

\subsection{Difference between NBQC and MB-DFTQC}
MBQC is a well-known model unique to quantum computing. Since typical MBQC repeatedly performs quantum teleportation of all algorithmic qubits to neighboring slices, it effectively removes the latency of entanglement generation from the critical path. In this sense, both MBQC and NBQC utilize the critical-path reduction techniques explained in the previous section. This section describes the qualitative differences between MBQC and NBQC and explains how circuit-specific NBQC designs can reduce the required number of nodes compared to MB-DFTQC.

\begin{figure*}[t]
    \centering
    \begin{minipage}{1.0\linewidth}
        \centering
        \includegraphics[width=\linewidth]{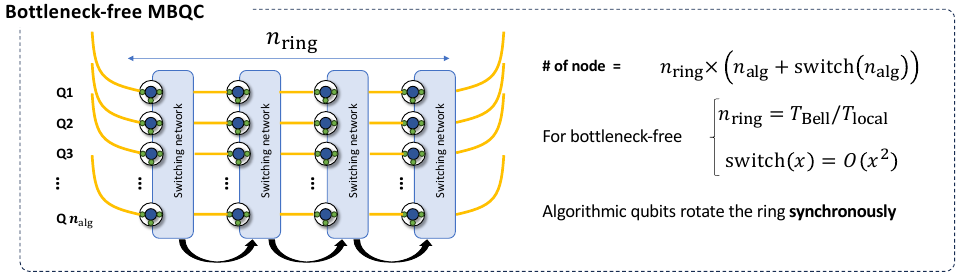}
        \subcaption{Example network design of measurement-based quantum computing~(MBQC). The algorithmic qubits stay on the same layer and synchronously teleport to the neighboring one. Each switching network needs to support all-to-all connectivity within a layer. This example shows the case with $n_{\rm ring}=4$.} 
        \label{fig:discuss_network_mbqc}
    \end{minipage}
    \begin{minipage}{1.0\linewidth}
        \centering
        \includegraphics[width=\linewidth]{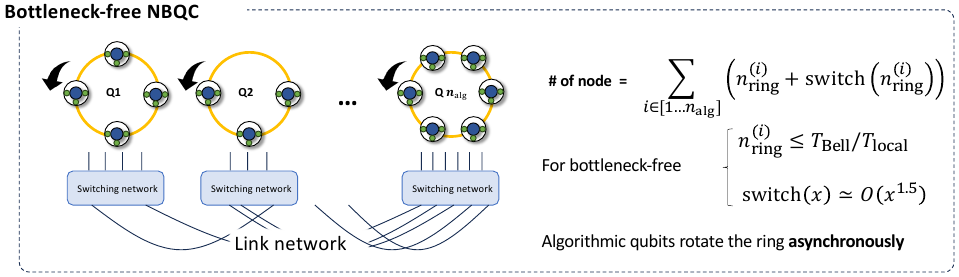}
        \subcaption{Example network design of network-based quantum computing~(NBQC). Each algorithmic qubit stays inside a ring network with an independent length and travels along the ring asynchronously. Each switching network needs to support a strict-sense non-blocking bipartite connection between the internal ports and external ports. This example shows the case with $n_{\rm ring}^{(1)}=4, n_{\rm ring}^{(2)}=3, \dots ,n_{\rm ring}^{(n_{\rm alg})}=6$.}
        \label{fig:discuss_network_nbqc}
    \end{minipage}
    \caption{The network design principle of measurement-based and network-based quantum computing.}
\end{figure*}

Figure~\ref{fig:discuss_network_mbqc} shows a typical network structure of MBQC. The network consists of a ring of $n_{\rm ring}$ layers, and each layer contains at least $n_{\rm alg}$ nodes if each node holds a single algorithmic qubit. We need to choose $n_{\rm ring}$ to be $O(T_{\text{Bell}}/T_{\text{local}})$ to ensure that communication latency is removed from the critical path. We also require additional nodes to construct a switching network to support flexible connectivity between algorithmic qubits. To enable all-to-all connectivity, the number of nodes per layer scales quadratically with the number of algorithmic qubits. Ignoring the cost of magic-state generation, the total number of nodes is given by the product of the number of layers and the number of nodes per layer. This design is similar to the circuit-agnostic NBQC, as it also synchronously rotates the positions of algorithmic qubits.

The network structure of circuit-specific NBQC, shown in Fig.~\ref{fig:discuss_network_nbqc}, consists of similar elements: a ring-shaped network and a switching network. However, there are two crucial differences from MBQC designs.
First, the ring network is separated, allowing us to choose non-uniform ring lengths and asynchronous movement of algorithmic qubits. This enables node-count efficient selection of ring lengths according to the access frequency of algorithmic qubits.
Second, the switching network is attached to each ring rather than to each layer. The node scaling of switching networks grows polynomially with the ring length, not the number of algorithmic qubits. In addition, the switching network needs to support bipartite communication between internal ports and external ports, not an all-to-all connection. Thus, the node count grows more slowly than the all-to-all connections.
Therefore, while both NBQC and MB-DFTQC can conceal communication latency and achieve algorithmic execution time based on similar principles, NBQC allows us to design a more efficient node count tailored to the target quantum circuits.

\subsection{Compatibility with code blocks containing multiple logical qubits}
Our benchmark focuses on the case where each node has a single algorithmic qubit with surface codes~(i.e., $n_{\rm node} = 1$). In practice, however, FTQC nodes may contain a larger number of $n_{\rm node}$, or one may employ QEC codes that encode multiple logical qubits within a single code block, such as quantum low-density parity-check codes~\cite{xu2024constant,bravyi2024high}.
Even in such cases, the NBQC principle can still be applied by treating the $n_{\rm node}$ data logical qubits as a bundled register that is teleported as a unit. To teleport $n_{\rm node}$ qubits, we need $n_{\rm node}$ entangled pairs to move them to neighboring nodes. This can be realized by performing entanglement distillation using the same QEC codes as the first-level QEC codes in the entanglement distillation process at the logical level~\cite{ataides2025constant}.
To make the NBQC network resource-efficient, the $n_{\rm alg}$ algorithmic qubits should be grouped into $n_{\rm node}$-sized registers, separating them into frequently accessed groups and rarely accessed ones. This optimization can be performed at the compilation stage or at runtime. We left the evaluation of these extensions and comparison to other frameworks as future work.

\subsection{Compatibility with devices beyond 2D connectivity}
The example implementation of fault-tolerant nodes shown in Fig.~\ref{fig:example_node_impl} assumes that each node can implement a 2D array of physical qubits. However, recent device technologies aim to realize qubit arrays beyond 2D connectivity by employing shuttling or long-range wiring~\cite{bluvstein2024logical,bravyi2024high,tremblay2022constant,berthusen2025toward,ueno2024high,sunami2025transversal}.
NBQC is compatible with such technologies because they only affect the list of available local operations and do not alter the resource-reduction principle of NBQC.
For instance, if neutral atoms and shuttling via optical tweezers~\cite{bluvstein2024logical,sunami2025transversal} are used, ancillary blocks~(gray cells) can be removed, $H$ gates can be performed via code automorphism, CNOT gates can be executed transversally, and $T/S$ gates can be implemented through magic-state teleportation.
Optimizing resource usage and comparing execution times across platforms would be an interesting direction for future work.

\subsection{Versatility of circuit-specific NBQC}
\label{subsec:discussion_specific_NBQC}
A major drawback of circuit-specific NBQC is that a network is specialized for a target quantum circuit. The bottleneck-free circuit-specific NBQC assumes the access-frequency profile of the target quantum circuit, and it is specialized to more details of target quantum circuits if the number of nodes is limited.
If we run quantum programs that have totally different profiles, the overhead of communication might become non-negligible. 
We can avoid this problem if we can reconfigure NBQC networks for each job, and we believe that this would be an acceptable cost for solving long-standing scientific problems in early regimes. On the other hand, network reconfiguration would become challenging as the number of nodes increases. 

Even if we cannot reconfigure the network, we believe circuit-specific NBQCs can show high performance if a target quantum circuit shares the common access patterns with a wide range of applications. 
While there are many types of quantum applications, they share a few base algorithms for exponential speed up, such as quantum phase estimation~\cite{babbush2018encoding}, quantum dynamics simulation~\cite{beverland2022assessing}, and quantum singular-value transformation~\cite{martyn2021grand}. Therefore, we can expect that applications that have the same base quantum algorithm would have a similar profile. For example, we can expect that programs of quantum phase estimation for different condensed-matter physics Hamiltonians would share a similar structure. Recently, it has been pointed out that typical quantum programs with exponential speed-up actually show characteristic memory-access patterns~\cite{kobori2025lsqca}. Exploring common profiles covering a wide range of applications and benchmarking them is the next step of this work.

We can also consider compiling a quantum program to fit an access profile expected by a given NBQC network. Algorithmic qubits in short and long ring networks can be regarded as slow and fast storage, respectively, and SWAP gates can be inserted to move frequently accessed qubits in subsequent operations into fast storage at runtime. Optimization techniques for similar constraints are developed in the state-vector simulations of quantum circuits in distributed classical computing~\cite{haner20175,imamura2022mpiqulacs}. In the full-state-vector simulation on distributed systems, qubits are classified into local qubits and global qubits, where local qubits do not demand communication between nodes while global ones do. To speed up the classical simulation, several SWAP gates are inserted to concentrate qubit accesses to local qubits. We can utilize these techniques to fit a target program to the expected access-frequency profile. Combining these characteristics of NBQC with compilation schemes to optimize the execution time is a promising direction for future work.

\section{Conclusion}
\label{sec:conclusion}
In this paper, we propose a framework for distributed fault-tolerant quantum computing, called Network-Based Quantum Computing~(NBQC), targeting a regime with many small FTQC nodes connected by slow quantum interconnects.
A key concept of NBQC is to conceal communication latencies using redundant nodes by performing repeated quantum teleportations over dedicated quantum networks.
NBQC achieves execution time similar to algorithmic execution time, i.e., a similar execution-time scaling to a single-node case. NBQC also enables the tunability between node count and execution time through iterative network updates.
We numerically evaluated the performance of NBQC and demonstrated that it can significantly reduce execution time with a modest increase in the number of nodes.
The NBQC framework is particularly effective when communication bottlenecks are localized to specific parts of the network.

There are several directions for further improving the performance of NBQC.
One important direction is co-design with entanglement distillation protocols. Since the rate of logical entanglement generation can be tuned according to the buffer size~\cite{pattison2025constant}, the number of computational nodes can be adaptively adjusted to minimize the total execution time.
Another direction is compile-time and runtime optimization dedicated to the NBQC architecture.
As shown in Table~\ref{tab:time_complexity}, NBQC exhibits favorable scaling when access to logical qubits is biased. Therefore, concentrating access on certain algorithmic qubits is effective, and this can be optimized either during the compilation phase or dynamically at runtime. Developing these technologies would open up a novel design space of distributed FTQCs.

\section*{Acknowledgments}
SN thanks Rodney Van Meter for the comment on the non-blocking switches. YS thanks Yosuke Ueno and Kae Nemoto for the comments on the contribution of this work from the perspective of computer architecture and distributed quantum computing. This work is supported by PRESTO JST Grant No.~JPMJPR1916, MEXT Q-LEAP Grant No.~JPMXS0120319794 and JPMXS0118068682, JST Moonshot R\&D Grant No.~JPMJMS2061, and JST CREST Grant No.~JPMJCR23I4 and  JPMJCR24I4.

\bibliography{reference}
\end{document}